\documentclass[journal]{IEEEtran}
\usepackage{multicol}
\usepackage{lipsum,graphicx}
\usepackage{float}
\usepackage{epstopdf}
\usepackage{ifthen}
\usepackage[latin1]{inputenc}
\usepackage{amssymb}
\usepackage[cmex10]{amsmath}
\usepackage{dcolumn}
\usepackage[nolist]{acronym}

\DeclareGraphicsExtensions{.pdf,.jpeg,.png,.eps}
\usepackage{booktabs}

\usepackage[caption=false,font=footnotesize]{subfig}
\usepackage{comment}
\usepackage{color}
\usepackage{cite}

\usepackage{setspace}
\usepackage{mathrsfs}          
\usepackage{amsmath, amssymb, amsthm}
\usepackage{amsfonts}
\usepackage{mathrsfs}
\usepackage{multirow}        
\usepackage{framed}
\usepackage{psfrag}
\usepackage{graphicx}
\usepackage{textcomp, gensymb} 
\usepackage{algpseudocode} 

\newcolumntype{d}[1]{D{.}{\cdot}{#1} }

\usepackage{bm}
\usepackage{mathtools}
\usepackage{siunitx}

\newcommand{\PreserveBackslash}[1]{\let\temp=\\#1\let\\=\temp}
\newcolumntype{C}[1]{>{\PreserveBackslash\centering}p{#1}}
\newcolumntype{R}[1]{>{\PreserveBackslash\raggedleft}p{#1}}
\newcolumntype{L}[1]{>{\PreserveBackslash\raggedright}p{#1}}

\newcommand{\figref}[1]{Fig.~\ref{#1}}

\newcommand{\tabref}[1]{Table~\ref{#1}}

\usepackage{lipsum}
\usepackage{mathtools}
\usepackage{cuted}
\usepackage{stmaryrd}

\usepackage{tabularx}
\usepackage{tabulary}
\usepackage{tikz}
\usepackage{pgfplots}
\usepgfplotslibrary{units}
\pgfplotsset{compat=1.18}

\usepackage{algpseudocode}
\usepackage{algorithm}
\usepackage{hyperref}

\begin{document}
	\begin{acronym}
\acro{1G}{first generation}
\acro{2G}{second generation}
\acro{3G}{third generation}
\acro{3GPP}{Third Generation Partnership Project}
\acro{4G}{fourth generation}
\acro{5G}{fifth generation}
\acro{802.11}{IEEE 802.11 specifications}
\acro{A/D}{analog-to-digital}
\acro{ADC}{analog-to-digital}
\acro{AM}{amplitude modulation}
\acro{AP}{access point}
\acro{AR}{augmented reality}
\acro{ASIC}{application-specific integrated circuit}
\acro{ASIP}{Application Specific Integrated Processors}
\acro{AWGN}{additive white Gaussian noise}
\acro{BCJR}{Bahl, Cocke, Jelinek and Raviv}
\acro{BER}{bit error rate}
\acro{BFDM}{bi-orthogonal frequency division multiplexing}
\acro{BPSK}{binary phase shift keying}
\acro{BS}{base stations}
\acro{CA}{carrier aggregation}
\acro{CAF}{cyclic autocorrelation function}
\acro{Car-2-x}{car-to-car and car-to-infrastructure communication}
\acro{CAZAC}{constant amplitude zero autocorrelation waveform}
\acro{CB-FMT}{cyclic block filtered multitone}
\acro{CCDF}{complementary cumulative density function}
\acro{CDF}{cumulative density function}
\acro{CDMA}{code-division multiple access}
\acro{CFO}{carrier frequency offset}
\acro{CIR}{channel impulse response}
\acro{CM}{complex multiplication}
\acro{COFDM}{coded-\acs{OFDM}}
\acro{CoMP}{coordinated multi point}
\acro{COQAM}{cyclic OQAM}
\acro{CP}{cyclic prefix}
\acro{CR}{cognitive radio}
\acro{CRC}{cyclic redundancy check}
\acro{CRLB}{Cram\'{e}r-Rao lower bound}
\acro{CS}{cyclic suffix}
\acro{CSI}{channel state information}
\acro{CSMA}{carrier-sense multiple access}
\acro{CWCU}{component-wise conditionally unbiased}
\acro{D/A}{digital-to-analog}
\acro{D2D}{device-to-device}
\acro{DAC}{digital-to-analog}
\acro{DC}{direct current}
\acro{DFE}{decision feedback equalizer}
\acro{DFT}{discrete Fourier transform}
\acro{DL}{downlink}
\acro{DMT}{discrete multitone}
\acro{DNN}{deep neural network}
\acro{DSA}{dynamic spectrum access}
\acro{DSL}{digital subscriber line}
\acro{DSP}{digital signal processor}
\acro{DTFT}{discrete-time Fourier transform}
\acro{DUT}{device under test}
\acro{DVB}{digital video broadcasting}
\acro{DVB-T}{terrestrial digital video broadcasting}
\acro{DWMT}{discrete wavelet multi tone}
\acro{DZT}{discrete Zak transform}
\acro{E2E}{end-to-end}
\acro{eNodeB}{evolved node b base station}
\acro{E-SNR}{effective signal-to-noise ratio}
\acro{EVD}{eigenvalue decomposition}
\acro{FBMC}{filter bank multicarrier}
\acro{FD}{frequency-domain}
\acro{FDD}{frequency-division duplexing}
\acro{FDE}{frequency domain equalization}
\acro{FDM}{frequency division multiplex}
\acro{FDMA}{frequency-division multiple access}
\acro{FEC}{forward error correction}
\acro{FER}{frame error rate}
\acro{FFT}{fast Fourier transform}
\acro{FIR}{finite impulse response}
\acro{FM}		{frequency modulation}
\acro{FMT}{filtered multi tone}
\acro{FO}{frequency offset}
\acro{F-OFDM}{filtered-\acs{OFDM}}
\acro{FPGA}{field programmable gate array}
\acro{FSC}{frequency selective channel}
\acro{FS-OQAM-GFDM}{frequency-shift OQAM-GFDM}
\acro{FT}{Fourier transform}
\acro{FTD}{fractional time delay}
\acro{FTN}{faster-than-Nyquist signaling}
\acro{GFDM}{generalized frequency division multiplexing}
\acro{GFDMA}{generalized frequency division multiple access}
\acro{GMC-CDM}{generalized	multicarrier code-division multiplexing}
\acro{GNSS}{global navigation satellite system}
\acro{GPS}{global positioning system}
\acro{GPSDO}{GPS disciplined oscillator}
\acro{GS}{guard symbols}
\acro{GSM}{Groupe Sp\'{e}cial Mobile}
\acro{GUI}{graphical user interface}
\acro{H2H}{human-to-human}
\acro{H2M}{human-to-machine}
\acro{HTC}{human type communication}
\acro{I}{in-phase}
\acro{i.i.d.}{independent and identically distributed}
\acro{IB}{in-band}
\acro{IBI}{inter-block interference}
\acro{IC}{interference cancellation}
\acro{ICI}{inter-carrier interference}
\acro{ICT}{information and communication technologies}
\acro{ICV}{information coefficient vector}
\acro{IDFT}{inverse discrete Fourier transform}
\acro{IDMA}{interleave division multiple access}
\acro{IEEE}{institute of electrical and electronics engineers}
\acro{IF}{intermediate frequency}
\acro{IFFT}{inverse fast Fourier transform}
\acro{IoT}{Internet of Things}
\acro{IOTA}{isotropic orthogonal transform algorithm}
\acro{IP}{internet protocole}
\acro{IP-core}{intellectual property core}
\acro{ISDB-T}{terrestrial integrated services digital broadcasting}
\acro{ISDN}{integrated services digital network}
\acro{ISI}{inter-symbol interference}
\acro{ITU}{International Telecommunication Union}
\acro{IUI}{inter-user interference}
\acro{LAN}{local area netwrok}
\acro{LLR}{log-likelihood ratio}
\acro{LMMSE}{linear minimum mean square error}
\acro{LNA}{low noise amplifier}
\acro{LO}{local oscillator}
\acro{LOS}{line-of-sight}
\acro{LoS}{line of sight}
\acro{LP}{low-pass}
\acro{LPF}{low-pass filter}
\acro{LS}{least squares}
\acro{LTE}{long term evolution}
\acro{LTE-A}{LTE-Advanced}
\acro{LTIV}{linear time invariant}
\acro{LTV}{linear time variant}
\acro{LUT}{lookup table}
\acro{M2M}{machine-to-machine}
\acro{MA}{multiple access}
\acro{MAC}{multiple access control}
\acro{MAP}{maximum a posteriori}
\acro{MC}{multicarrier}
\acro{MCA}{multicarrier access}
\acro{MCM}{multicarrier modulation}
\acro{MCS}{modulation coding scheme}
\acro{MF}{matched filter}
\acro{MF-SIC}{matched filter with successive interference cancellation}
\acro{MIMO}{multiple-input multiple-output}
\acro{MISO}{multiple-input single-output}
\acro{ML}{machien learning}
\acro{MLD}{maximum likelihood detection}
\acro{MLE}{maximum likelihood estimator}
\acro{MMSE}{minimum mean squared error}
\acro{MRC}{maximum ratio combining}
\acro{MS}{mobile stations}
\acro{MSE}{mean squared error}
\acro{MSK}{Minimum-shift keying}
\acro{MSSS}[MSSS]	{mean-square signal separation}
\acro{MTC}{machine type communication}
\acro{MU}{multi user}
\acro{MVUE}{minimum variance unbiased estimator}
\acro{NEF}{noise enhancement factor}
\acro{NLOS}{non-line-of-sight}
\acro{NMSE}{normalized mean-squared error}
\acro{NOMA}{non-orthogonal multiple access}
\acro{NPR}{near-perfect reconstruction}
\acro{NRZ}{non-return-to-zero}
\acro{OFDM}{orthogonal frequency division multiplexing}
\acro{OFDMA}{orthogonal frequency division multiple access}
\acro{OOB}{out-of-band}
\acro{OQAM}{offset quadrature amplitude modulation}
\acro{OQPSK}{offset quadrature phase shift keying}
\acro{OTFS}{orthogonal time frequency space}
\acro{PA}{power amplifier}
\acro{PAM}{pulse amplitude modulation}
\acro{PAPR}{peak-to-average power ratio}
\acro{PC-CC}{parallel concatenated convolutional code}
\acro{PCP}{pseudo-circular pre/post-amble}
\acro{PD}{probability of detection}
\acro{pdf}{probability density function}
\acro{PDF}{probability distribution function}
\acro{PDP}{power delay profile}
\acro{PFA}{probability of false alarm}
\acro{PHY}{physical layer}
\acro{PIC}{parallel interference cancellation}
\acro{PLC}{power line communication}
\acro{PMF}{probability mass function}
\acro{PN}{pseudo noise}
\acro{ppm}{parts per million}
\acro{PPS}{pulse per second}
\acro{PRB}{physical resource block}
\acro{PRB}{physical resource block}
\acro{PSD}{power spectral density}
\acro{Q}{quadrature-phase}
\acro{QAM}{quadrature amplitude modulation}
\acro{QoS}{quality of service}
\acro{QPSK}{quadrature phase shift keying}
\acro{R/W}{read-or-write}
\acro{RAM}{random-access memmory}
\acro{RAN}{radio access network}
\acro{RAT}{radio access technologies}
\acro{RC}{raised cosine}
\acro{RF}{radio frequency}
\acro{rms}{root mean square}
\acro{RRC}{root raised cosine}
\acro{RW}{read-and-write}
\acro{SC}{single-carrier}
\acro{SCA}{single-carrier access}
\acro{SC-FDE}{single-carrier with frequency domain equalization}
\acro{SC-FDM}{single-carrier frequency division multiplexing}
\acro{SC-FDMA}{single-carrier frequency division multiple access}
\acro{SD}{sphere decoding}
\acro{SDD}{space-division duplexing}
\acro{SDMA}{space division multiple access}
\acro{SDR}{software-defined radio}
\acro{SDW}{software-defined waveform}
\acro{SEFDM}{spectrally efficient frequency division multiplexing}
\acro{SE-FDM}{spectrally efficient frequency division multiplexing}
\acro{SER}{symbol error rate}
\acro{SIC}{successive interference cancellation}
\acro{SINR}{signal-to-interference-plus-noise ratio}
\acro{SIR}{signal-to-interference ratio}
\acro{SISO}{single-input, single-output}
\acro{SMS}{Short Message Service}
\acro{SNR}{signal-to-noise ratio}
\acro{STC}{space-time coding}
\acro{STFT}{short-time Fourier transform}
\acro{STO}{symbol time offset}
\acro{SU}{single user}
\acro{SVD}{singular value decomposition}
\acro{TD}{time-domain}	
\acro{TDD}{time-division duplexing}
\acro{TDMA}{time-division multiple access}
\acro{TFL}{time-frequency localization}
\acro{TO}{time offset}
\acro{TS-OQAM-GFDM}{time-shifted OQAM-GFDM}
\acro{UE}{user equipment}
\acro{UFMC}{universally filtered multicarrier}
\acro{UL}{uplink}
\acro{US}{uncorrelated scattering}
\acro{USB}{universal serial bus}
\acro{UW}{unique word}
\acro{VLC}{visible light communications}
\acro{VR}{virtual reality}
\acro{WCP}{windowing and \acs{CP}}	
\acro{WHT}{Walsh-Hadamard transform}
\acro{WiMAX}{worldwide interoperability for microwave access}
\acro{WLAN}{wireless local area network}
\acro{W-OFDM}{windowed-\acs{OFDM}}	
\acro{WOLA}{windowing and overlapping}	
\acro{WSS}{wide-sense stationary}
\acro{ZCT}{Zadoff-Chu transform}
\acro{ZF}{zero-forcing}
\acro{ZMCSCG}{zero-mean circularly-symmetric complex Gaussian}
\acro{ZP}{zero-padding}
\acro{ZT}{zero-tail}
\acro{URLLC}{ultra-reliable low-latency communications}
\acro{PLL}{phase-locked loop}
\acro{USRP}{universal software radio peripheral}
\acro{TX}{transmission}
\acro{REF}{reference}
\acro{PFD}{phase frequency detector}
\acro{LF}{loop filter}
\acro{VCO}{voltage-controlled oscillator}
\acro{TIE}{time interval error}
\acro{ACF}{autocorrelation function}
\acro{OU}{Ornstein-Uhlenbeck}
\acro{CCF}{cross-correlation function}
\acro{WSS}{wide-sense stationary}
\acro{SDE}{stochastic differential equation}

\acro{HSI}{human system interface}
\acro{HMI}{human machine interface}
\acro{VR} {visual reality} 
\acro{AGV}{automated guided vehicles}
\acro{MEC}{multiaccess edge cloud}
\acro{TI} {tactile Internet}
\acro{IMT}{ international mobile telecommunications}
\acro{GN}{gateway node}
\acro{CN}{control node}
\acro{NC}{network controller}
\acro{SN}{sensor node}
\acro{AN}{actuator node}
\acro{HN}{haptic node}
\acro{TD}{tactile devices}
\acro{SE}{supporting engine}
\acro{AI}{artificial intelligence}
\acro{TSM}{tactile service manager}
\acro{TTI}{transmission time interval}
\acro{NR}{new radio}
\acro{SDN}{software defined networking}
\acro{NFV}{ network function virtualization}
\acro{CPS}{cyber-physical system}
\acro{TSN}{Time-Sensitive Networking}
\acro{FEC}{forward error correction}
\acro{STC}{space-time  coding}
\acro{HARQ}{hybrid automatic repeat request}
\acro{CoMP} {Coordinated multipoint}
\acro{HIS}{human system interface }
\acro{RU}{radio unit}
\acro{CU}{central unit}
\acro{AoD} {angle of departure}
\end{acronym}
	\title{A Practical Analysis: Understanding Phase Noise Modelling in Time and Frequency Domain for Phase-Locked Loops} 
	
	\author{
		\IEEEauthorblockN{
			Carl Collmann, Bitan Banerjee, Ahmad Nimr, Gerhard Fettweis
			}
			
		\IEEEauthorblockA{
		Vodafone Chair Mobile Communications Systems, Technische Universit\"{a}t Dresden, Germany\\ \small\texttt{\{carl.collmann, bitan.banerjee, ahmad.nimr,  gerhard.fettweis\}@tu-dresden.de}\\
		}
		}
	\maketitle
	\IEEEpeerreviewmaketitle
	\begin{abstract}

In \ac{MIMO} systems, the presence of phase noise is a significant factor that can degrade performance.
For \ac{MIMO} testbeds build from \ac{SDR} devices, phase noise cannot be ignored, particular in applications that require phase synchronization.
This is especially relevant in  \ac{MIMO} systems that employ digital beamforming, where precise phase alignment is crucial.
Accordingly, accurate phase noise modelling of \ac{SDR} devices is essential.
However, the information provided in data sheets for different \ac{SDR} models varies widely and is often insufficient for comprehensive characterization of their phase noise performance.
While numerical simulations of \ac{PLL} phase noise behavior are documented in the literature, there is a lack of extensive measurements supported by appropriate system modelling.
In this work, we present a practical phase noise modeling methodology applied to an \ac{SDR} from the \ac{USRP} \texttt{X310} series.
Based on measurement data, we derive estimates of key \ac{PLL} performance indicators such as cycle-to-cycle jitter, oscillator constants, and \ac{PLL} bandwidth.
Furthermore, we propose a parametric model for the phase noise \ac{PSD} of the \ac{PLL} circuit and provide corresponding parameter estimates.
This model can be used for further investigation into the impact of phase noise on \ac{MIMO} system performance implemented by similar \ac{SDR} devices.
\end{abstract}


	\acresetall
\section{Introduction}\label{sec:introduction}

The upcoming sixth generation (6G) of mobile communication networks aims to achieve improved data rates, low latency, and highly reliable connections \cite{6gvision23}.
To meet these specifications, increased bandwidth at higher carrier frequency is adopted.
However, at higher frequencies and bandwidths, the effect of hardware impairments such as phase noise, becomes significantly more pronounced and cannot be ignored.
Therefore, it is essential to develop rigorous models and thorough explanations of phase noise effect.

In joint communications and sensing systems (JC\&S), the common phase error (CPE), to which phase noise contributes, can severely influence the system performance.
The phase noise causes intercarrier interference (ICI), which reduces the efficiency of data transmission.
Additionally, the sensing accuracy can be adversely affected by the presence of phase noise \cite{bsbf21}.
Similarly, in \ac{MIMO} systems, phase noise can degrade spatial multiplexing gains because it deteriorates the coherent combination of signals \cite{Heath_mimo_18}.

A reliable model of phase noise is required to enable the design of algorithms capable of mitigating its impairing effects.
While the topic of phase noise in frequency-synthesizing circuits has received significant attention, many existing works face limitations.
For instance, in \cite{demir847872} and \cite{mehrotra1031966}, a rigorous unifying theory of phase noise analysis is introduced, which has gained widespread acceptance.
However, this theory is presented without explicit validation through measurements.
Subsequent works, such as \cite{loehning8257361325},  build upon this theory and improve its accessibility but still fail to establish strong links between theory and empirical observations.
In \cite{david_radu7119074}, a method for phase tracking using a Kalman filter is investigated and experimentally validated with \ac{SDR}.
However, the authors investigate primarily the Allan deviation rather  than explicitly addressing the phase noise process.
Other works, such as \cite{tschapek9721401}, primarily focus on presenting phase noise measurements in different scenarios but provide limited corresponding modelling of the phase noise.

The 3GPP standard also offers a phase noise model \cite{3gppTR}. 
For a specific \ac{PLL},  measurements are first performed, and then the parameters of a digital filter are fitted to these measurements to generate an empirical phase noise model. However, estimators for these parameters are not provided, and the  model is limited to the phase noise spectrum.
Consequently, the time-domain behavior of the phase noise process, particularly its non-stationary characteristics, can not be extrapolated.
Furthermore, the dynamics of the \ac{PLL} circuit are not modeled.
While this kind of analysis can provide useful abstraction, more dedicated approaches are known \cite{demir847872} and might be more applicable, depending on the desired use-case of the model.


\subsection{Contribution}
The contributions of this paper are as follows
\begin{itemize}
    \item A simplified parametric phase noise spectrum model is derived by modelling the phase noise process and \ac{PLL} dynamics in the time domain.
    \item The phase noise spectrum for common \ac{SDR} devices is measured, and the impact of various \ac{RF} frontend daughterboards and frequencies is analyzed.
    \item Estimators for the parameters of the derived phase noise model are provided, bridging  the established theoretical model with empirical observation.
\end{itemize}

\subsection{Organization}
The rest of the paper is organized as follows: In section \ref{sec:syst_model}, a general system model for phase noise is described.
Next, a model for a free-running \ac{VCO} in the time domain and its corresponding phase noise spectrum is derived in section \ref{sec:vco}. This model is then expanded to a \ac{PLL} circuit in section \ref{sec:pll}.
A model of stochastic differential equations in the time domain is presented along with the corresponding derivations to obtain the phase noise spectrum.
In section \ref{sec:measurement}, the measurement setup is introduced.
An overview of the recorded datasets are provided, highlighting the distinction of the phase noise spectrum of different daughterboard models for the \ac{USRP}'s evaluated.
In section \ref{sec:parameter}, the previously derived model is expanded to match the characteristics of the \ac{DUT}.
Estimators for deriving system parameters are introduced, and the corresponding  estimates are presented for two \ac{USRP} daughterboard models.Finally, the paper is concluded in section \ref{sec:conclusion}.

	\section{Overview of Phase Noise}\label{sec:syst_model} 

Phase noise generally refers to a random change in the phase of a signal.
In the time domain, this random change is commonly referred to as jitter. 
IEEE \cite{IEEE4797525} defines phase noise as the single-sideband phase noise spectrum of the \ac{PSD} of phase fluctuations.
Phase noise is given as $\mathcal{L}(f_\text{offs})$ in $\SI{}{dBc/Hz}$ (power relative to carrier) at offset $f_\text{offs}$ relative to carrier frequency.
While an ideal oscillator produces a signal of a constant frequency, real oscillators are not stable in frequency.
This fluctuation in frequency is linked to a corresponding \ac{TIE} or fluctuation in the time domain.
To characterize this fluctuation, we use a stochastic time shift $\alpha(t)$.
The time shift can be translated into a phase shift $\phi(t)=2\pi f_0 \alpha(t)$ for a given oscillator frequency.\\

In the following, the effect of a stationary phase noise process in the time domain for a free running oscillator is analyzed and properties of a non-stationary phase noise process are described.
This model is extended in following sections, where at first the impact of non-stationary phase noise processes on free running oscillators is explored.
Finally, the model is extended to the entire \ac{PLL} circuit.
Each model covers time domain signals of phase noise processes and the corresponding frequency domain representations.
The presented derivations follow those established in previous research \cite{demir847872,mehrotra1031966,loehning8257361325}. 

\subsection{PSD of Oscillators Affected by a Stationary Noise Process} 

\begin{figure}[htb]
    \centering
    \includegraphics[width=0.8\linewidth]{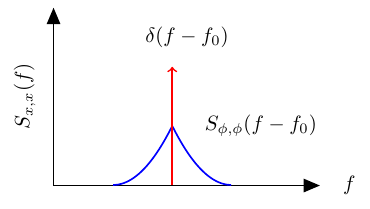}
    \caption{\ac{PSD} of a generic noisy oscillator with phase noise spectrum $S_{\phi,\phi}$}
    \label{fig:generic_osc_model}
\end{figure}

The output signal of an ideal oscillators first harmonic can be described as 
\begin{align}
    x_0 (t) = A ~ x_\text{CLK} (2\pi f_0 t + \phi_0).
    \label{eq:ideal_osc}
\end{align}
The amplitude of the oscillator signal $A$ is assumed to be constant. 
The default phase shift of the oscillator signal is called $\phi_0$.
The term $x_\text{CLK}$ represents a generic oscillator clock function such as sine or square wave.
In the following, $\cos$ is used as oscillator clock function, with default frequency $f_0$.
The output signal of a disturbed oscillator, applying the mentioned simplifications, is
\begin{align}
    x_\text{PN} (t) =  \cos (2\pi f_0 t + \phi (t)) = \mathfrak{Re} \left\{ e^{j ( \omega_0 t + \phi (t))} \right\}.
    \label{eq:dist_osc}
\end{align}
The time dependent phase shift process $\phi (t)$ representing the phase noise, can also be modeled by a time shift process $\alpha(t) = \phi (t) / 2\pi f_0$.
In this section, $\phi (t)$ is assumed to be a stationary so that the \ac{PSD} can be computed.
To simplify the following derivation, the complex representation of the oscillator signal is used
\begin{align}
    x_\text{PN} = e^{j(\omega_0 t + \phi (t))} = e^{j(\omega_0 t + \omega_0 \alpha (t))}.
\end{align}
Assuming that the term $\phi (t)$ is sufficiently small, a Taylor series approximation is applied
\begin{align}
    x_\text{PN} \approx e^{j\omega_0 t} \left( 1 + j \phi (t) \right).
\end{align}
Next, the \ac{ACF} is calculated, assuming that the \ac{ACF} of the noise term $ R_{\phi,\phi}(\tau)$ is known. 
Furthermore it is assumed that the oscillator signal is uncorrelated with the zero mean noise process so that the corresponding cross terms in the expectation vanish, 
\begin{equation}
   \begin{split}
        R_{x,x}(t, t + \tau) &= \mathbb{E} \left[ x_\text{PN}(t)x_\text{PN}^{*}(t + \tau) \right]\\
        &= e^{-j\omega_0 \tau} \left( 1 + R_{\phi,\phi}(\tau) \right)
  \end{split}
\end{equation}
To obtain the \ac{PSD}, the Fourier transform of the \ac{ACF} has to be calculated.
In this context $S_{\phi,\phi}(f) = \mathfrak{F}_\tau \Bigl\{ R_{\phi,\phi}(\tau) \Bigr\}$ denotes the phase noise \ac{PSD}.
Then the output spectrum of a generic noisy oscillator is given as 
\begin{align}
    S_{x,x}(f) &= \mathfrak{F}_\tau \Bigl\{ R_{x,x} (t, t + \tau) \Bigr\}\nonumber\\
    &=\delta(f - f_0) + S_{\phi,\phi}(f - f_0)
\end{align}

It can be seen that the \ac{PSD} consists of a Dirac pulse at the oscillators frequency and the noise spectrum component that centered around the oscillator frequency $f_0$ (as illustrated in Fig. \ref{fig:generic_osc_model})

\subsection{Wiener Process} 

In established research, it is common practice to model the phase noise as a Wiener process \cite{demir847872}.
A Wiener process is a non-stationary stochastic process characterized by independent normally distributed increments.
The Wiener process $W(t)$ can be characterized as the integral over a zero mean, white Gaussian process $\xi(t)$ of unit variance
\begin{align}
              W(t) &= \int_{0}^{t} \xi(t')\, dt'
              = \int_{0}^{W(t)}\, dW(t')\label{eq:diff_wiener}.
\end{align}
Note that a differential step of a Wiener process is defined as
\begin{align}
    dW(t) \equiv W(t + dt) - W(t) = \xi(t)dt. \label{eq:wiener_process_diff}
\end{align}

It is assumed that the stochastic time shift ${\alpha(t) = \sqrt{c_\text{VCO}} W(t)}$ affecting a free running oscillator can be modelled as a Wiener process.
In this case, the Wiener process is scaled with the oscillator specific constant $c_\text{VCO}$.
Then, the following properties hold for the process of stochastic time shifts \cite{higham0036144500378302}, \cite[p. 445 ff.]{papoulis_pillai78965125}:
\begin{itemize}
    \item[1.] $\alpha(t)$ is zero mean and not stationary
    \begin{align*}
        \mathbb{E} \left[ \alpha(t) \right] = 0.
    \end{align*}
    \item[2.] The variance of the process grows linear with time $t$
    \begin{align}
        \sigma_\alpha^{2}(t) = \mathbb{E} \left[|\alpha(t)|^2 \right] = c_\text{VCO} t \label{eq:var_alpha}.
    \end{align}
    \item[3.] At time instance $\tau$, the random variable $\alpha(t=\tau)$ is gauss-distributed.
    Its density function is given as 
    \begin{align*}
        f_\alpha(z,t) = \frac{1}{\sqrt{2\pi c_\text{VCO} t}} e^{-\frac{z^2}{2 c_\text{VCO} t}}.
    \end{align*}
    \item[4.] increments of the process $\alpha(t)$ are independent
    \begin{align*}
        \mathbb{E} = \left[ (\alpha(\tau_1) - \alpha(\tau_0))(\alpha(\tau_0)) \right] = 0,~ \text{for}~\tau_0 < \tau_1.
    \end{align*}
    \item[5.] The autocorrelation function for the process $\alpha(t)$ is given as (c.f. \cite[Corollary 7.1]{demir847872})
    \begin{align}
        R_{\alpha,\alpha} (t, t + \tau) 
        &= c_\text{VCO} t, ~ \text{for}~ \tau > 0 . \label{eq:acf_alpha}
    \end{align}
\end{itemize}

	\section{Free Running Oscillator}\label{sec:vco}

This section describes the modeling of a generic free-running oscillator affected by a Wiener phase noise.
The model considers a \ac{VCO} subjected to white Gaussian noise at its control voltage.
Key findings from prior analyses \cite{demir847872} are: 
\begin{itemize}
    \item[1.] The process $\alpha(t)$ of random time shifts affecting the ideal free running oscillator can asymptotically, i.e.,  $t \rightarrow \infty$, be described as a Wiener-process and is Gaussian distributed.
    \item[2.] The output of a noisy oscillator is asymptotically stationary. Its corresponding \ac{PSD} consists of Lorentzian spectra, symmetrically centered around the harmonics of an undisturbed oscillator.
    \item[3.] A scalar constant $c_\text{VCO}$ is sufficient to characterize the process $\alpha(t)$ and the corresponding \ac{PSD} of a noisy oscillator.
\end{itemize}
\subsection{Voltage Controlled Oscillator Model}
Analyzing the case of a free-running \ac{VCO}, further assumptions are:
\begin{itemize}
    \item[4.] The oscillator is affected by independent white noise processes that can be combined to a single white Gaussian noise source at the input of the \ac{VCO}.
    \item[5.] A shift in the control voltage of the oscillator at the input $\Delta U$ corresponds linearly to a frequency shift $\Delta f$. The voltage sensitivity $K_\text{VCO}$ in $[\SI{}{Hz/V}]$ is an oscillator-specific constant.
    The frequency of the oscillator output signal can then be expressed as
    \begin{align}
        f &= f_0 + \underbrace{K_\text{VCO} \Delta U(t)}_{\Delta f(t)}.
    \end{align}        
\end{itemize}
The output voltage signal can be described using the stochastic process $\Delta f(t)$ representing the instantaneous frequency change as follows, 
\begin{align}
    x_\text{VCO}(t) = \cos \left( 2\pi f_0 t  + \underbrace{2\pi \int_{0}^{t} \Delta f(\tau)\, d\tau}_{ \phi(t)} + \phi_0  \right),
\end{align}
with $\phi(t)$ referring to the phase shift process.
The non-idealities of the noisy oscillator are characterized with regards to a stochastic time shift $\alpha(t)$.
This time shift is connected to the phase shift and frequency shift such that
\begin{align}
    \alpha(t) &= \frac{1}{2\pi f_0} \phi(t) = \frac{1}{f_0} \int_{0}^{t} \Delta f(t')\, dt'.
\end{align}

\subsection{Free Running Oscillator - Discrete Time Domain Model}

The model described here represents a free running oscillator in discrete time.
As established previously, the stochastic time shift process of the oscillator can be modeled as a Wiener process.
This process is a Gaussian process $\xi$ of unit variance, scaled with the oscillator constant $c_\text{VCO}$.
In discrete time, the process can be written as
\begin{align}
    \alpha_\text{VCO}[n] = 
    \begin{cases}
      \xi(0) = 0, & n = 0 \\
      \sqrt{c_\text{VCO} \Delta t} \sum_{i=0}^{n-1} \xi(i \Delta t), & n > 0
    \end{cases}.
    \label{eq:pn_vco}
\end{align}
Defining $\xi(0) = 0$ is a convention and has no significance on the statistical properties of the time shift process.
The term $\xi(i \Delta t)$ refers to increments at discrete time points $i\Delta t$, with $\Delta t$ representing the sample interval. 
The corresponding phase shift process is then given as $\phi_\text{VCO}[n] = 2 \pi f_0 \alpha_\text{VCO}[n]$.

\subsection{Autocorrelation Function of Voltage Controlled Oscillator Output Process}

\begin{figure}[t]
    \centering
    \includegraphics[width=0.7\linewidth]{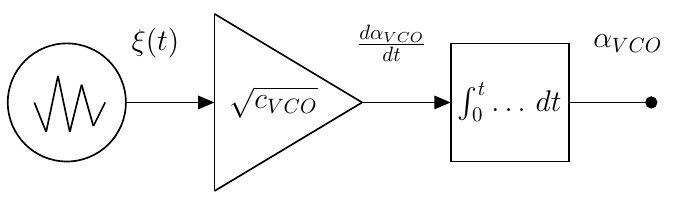}
    \caption{Noise equivalent circuit diagram for a \ac{VCO}, noise process $\xi(t)$ from noise source input, scaled with $\sqrt{c_\text{VCO}}$, integral over differential time shifts yields process $\alpha_\text{VCO}$}
    \label{fig:noise_equivalent_circuit}
\end{figure}

In this subsection, the aim is to find an analytic expression for the \ac{ACF} at the output of the noisy oscillator
This \ac{ACF} can then be used to compute the \ac{PSD} in the next step.
The noise equivalent circuit diagram for a \ac{VCO} is given in Fig. \ref{fig:noise_equivalent_circuit}.
An ideal oscillator signal (cf. \eqref{eq:ideal_osc}) is a periodic signal that can be expressed by the Fourier series
\begin{align}
    x_0(t) = \sum_{i=-\infty}^{\infty} \underline{X}_i e^{j i 2\pi f_0 t}, \label{eq:fourier_series_ideal_osc}
\end{align}
with complex conjugate symmetric Fourier coefficients $\underline{X}_i$.
The \ac{ACF} of the noisy oscillator output for the first harmonic $i=\pm 1$ is then given as 
\begin{align}
\begin{split}
    R_{x,x} &(t, t + \tau) = \mathbb{E} \left[ x_0(t+\alpha(t))x_0^{*}(t + \tau + \alpha(t + \tau)) \right]\nonumber
\end{split} \\
\begin{split}
    &= \sum_{i=\pm 1} {\left| \underline{X}_i \right|^2 e^{ji2\pi f_0 \tau} \overbrace{\mathbb{E} \left[ e^{ji 2\pi f_0 [\alpha(t) - \alpha(t+\tau)]}\right]}^{\Psi_{\Delta\alpha}(i 2\pi f_0)}} \\
    &+ {\underline{X}_i \underline{X}_{-i}^{*} e^{ji 2\pi f_0 (2t + \tau)} \overbrace{\mathbb{E} \left[ e^{ji 2\pi f_0 [\alpha(t) + \alpha(t + \tau)]} \right]}^{\Psi_{\Sigma\alpha}(i  2\pi f_0)}}.\label{eq:acf}
\end{split}
\end{align}
In the expression in \eqref{eq:acf}, there is a term for a difference process $\Delta\alpha = \alpha(t) -\alpha(t + \tau)$ and a sum process $\Sigma\alpha=\alpha(t) + \alpha(t + \tau)$. 
The corresponding characteristic functions\footnote{characteristic function $\Psi_Z(t)=\mathbb{E} \left[ e^{jtZ}\right]$ of Gaussian-distributed random variable $Z$} are specified to replace the terms of the expectation \cite{papoulis_pillai78965125}
\begin{align}
    \Psi_{\Delta\alpha}(i  2\pi f_0) &= e^{-\frac{1}{2} (2\pi f_0)^2 \text{Var}(\Delta\alpha)} \label{eq:psi_delta},\\
    \Psi_{\Sigma\alpha}(i  2\pi f_0) &= e^{-\frac{1}{2} (2\pi f_0)^2 \text{Var}(\Sigma\alpha)}.\label{eq:psi_sigma}
\end{align}
Using properties of the wiener process, the variance is found (c.f. \cite[(32)]{demir847872} with $i=k,i\neq k$ respectively, assuming $\tau > 0$)
\begin{align}
    \text{Var}(\Delta\alpha) &= \text{Var}(\alpha(t) - \alpha(t + \tau)), 
    = c_\text{VCO} | \tau |\label{eq:var_diff}
\end{align}
and for the sum process
\begin{align}
    \text{Var}(\Sigma\alpha) &= \text{Var}(\alpha(t) + \alpha(t + \tau)) 
    = c_\text{VCO} \left[ 4t + \tau \right].\label{eq:var_sum}
\end{align}
Note that the characteristic function for the sum process $\Psi_{\Sigma\alpha}$ becomes asymptotically  ${\lim\limits_{t\to\infty} e^{-t} = 0}$.
However, the characteristic function for the difference term $\Psi_{\Delta\alpha}$ is independent of time.
Then, the \ac{ACF} asymptotic time invariant behaviour for $t \rightarrow \infty$ can be obtained by inserting into (\ref{eq:acf}).
\begin{align}
\lim_{t\to\infty}  R_{x,x} (t, t + \tau) &=  \sum_{i=\pm 1} \left| \underline{X}_i \right|^2 e^{ji 2\pi f_0 \tau} e^{-\frac{1}{2} (2\pi f_0)^2 c_\text{VCO} | \tau |} \nonumber\\
&=  R_{x,x} (\tau).
\end{align}
\subsection{Power Spectrum Density of Voltage Controlled Oscillator}
\begin{figure}[t]
    \centering
    \includegraphics[width=0.8\linewidth]{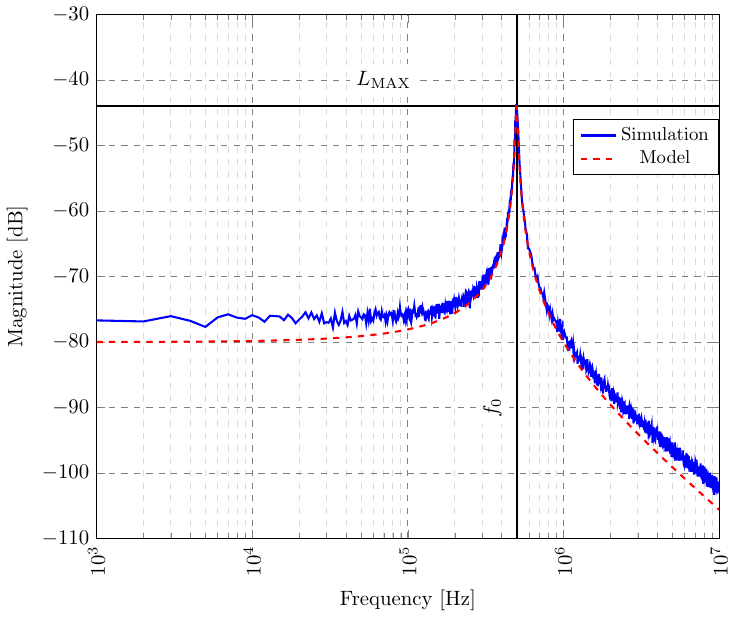}
    \caption{Noisy oscillator output magnitude spectrum $\left|S_{x,x}(f)\right|^2$ around first harmonic, $f_0 = f_\text{VCO} = \SI{500}{kHz}$, $f_s = \SI{100}{MHz}$ sampling frequency, $N=\SI{100}{k}$ samples, $c_\text{VCO} = \SI{e-11}{s}$ oscillator constant}
    \label{fig:phase_noise_figure_f}
\end{figure}

\begin{figure}[t]
    \centering
    \includegraphics[width=0.8\linewidth]{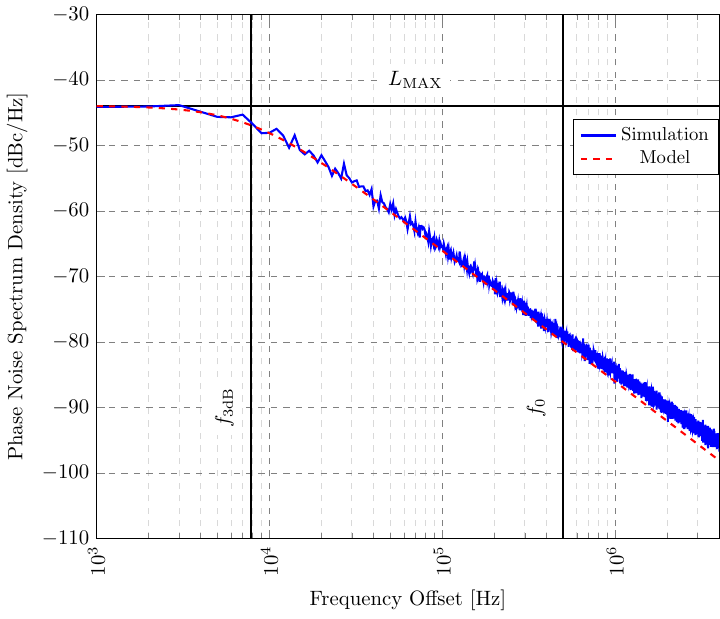}
    \caption{Phase noise spectrum, $f_0 = f_\text{VCO} = \SI{500}{kHz}$, $f_s = \SI{100}{MHz}$ sampling frequency, $N=\SI{100}{k}$ samples, $c_\text{VCO} = \SI{e-11}{s}$ oscillator constant}
    \label{fig:phase_noise_figure_f_offs}
\end{figure}

The \ac{PSD} of the disturbed oscillator output is computed by the Fourier transform $ S_{x,x}(f) = \mathfrak{F}_\tau \Bigl\{ R_{x,x} (\tau) \Bigr\}$, and given by 
\begin{align}
    S_{x,x}(f) &= \sum_{i=\pm 1} \left| \underline{X}_i \right|^2 \frac{f_\text{0}^2 c_\text{VCO}}{(f + if_\text{0})^2 + \pi^2f_\text{0}^4 c_\text{VCO}^2}.\label{eq:phase_noise_psd_omega}
\end{align}
The \ac{PSD} shape of the oscillator output is a Lorentzian, shifted by the oscillator frequency $f_0$ (cf. noisy oscillator output spectrum around first harmonic, Fig. \ref{fig:phase_noise_figure_f}).
The phase noise of the oscillator has the effect of spreading the spectrum of the dirac pulse which represents the \ac{PSD} of ideal oscillator.
The spectral spreading is characterized by $c_\text{VCO}$ of the oscillator.
To obtain the single-sideband phase noise spectrum $\mathfrak{L}(f_\text{offs})$, (\ref{eq:phase_noise_psd_omega}) is normed to the power at the first harmonic with $f_\text{offs}$ frequency offset to carrier (see Fig. \ref{fig:phase_noise_figure_f_offs})
\begin{align}
    \mathfrak{L}(f_\text{offs}) &= 10 \log_{10} \left( \frac{S_{x,x}(f_0 + f_\text{offs})}{\sum_{i=\pm 1} \left| \underline{X}_i \right|^2} \right)\nonumber\\
                           &= 10 \log_{10} \left( \frac{f_0^2 c_\text{VCO}}{f_\text{offs}^2 + \pi^2 f_0^4 c_\text{VCO}^2} \SI{1}{Hz}\right) \SI{}{dBc/Hz}.\label{eq:phase_noise_psd_f_offs}
\end{align}
The maximum \ac{PSD} component can be found where the offset frequency becomes zero, 
\begin{align}
    \mathfrak{L}_{MAX} &= -10 \log_{10} \left(\pi^2 f_0^2 c_\text{VCO}\right).
    \label{eq:l_max}
\end{align}
In Fig.~\ref{fig:phase_noise_figure_f_offs}, the low-pass characteristic of the phase noise \ac{PSD} can be observed.
The term $f_{3dB}=\pi f_0^2 c_\text{VCO}$ refers to the $\SI{3}{dB}$ cut-off frequency.
In the literature, this cut-off frequency is also known as \textit{phase noise bandwidth} \cite{Petrovic2003PhaseNS}.
	\begin{figure*}[htb]
		\centering
    \includegraphics[width=0.8\textwidth]{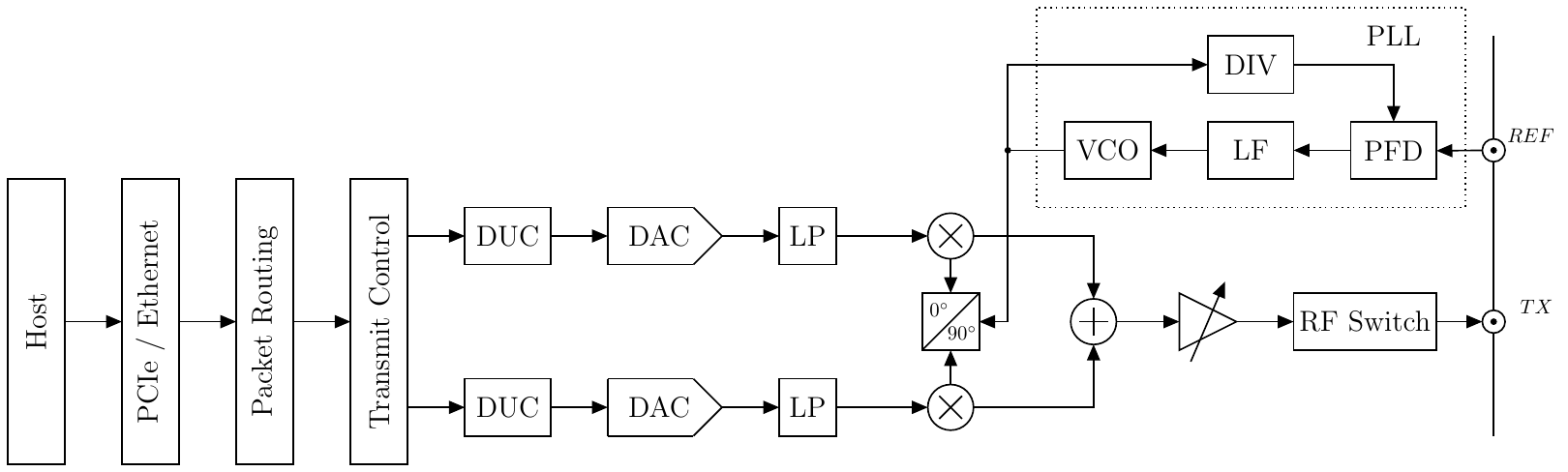}
       \caption{Block diagram of the \ac{TX} \ac{RF} Chain for a generic \ac{USRP}, \ac{PLL} receives input from \ac{REF} port, the \ac{PLL} consists of \ac{PFD}, \ac{LF}, \ac{VCO} and a frequency divider}
	    \label{fig:usrp2953_block_diagram}
\end{figure*}

\section{Phase-Locked Loop} \label{sec:pll}

\begin{figure}[b]
    \centering
    \includegraphics[width=\linewidth]{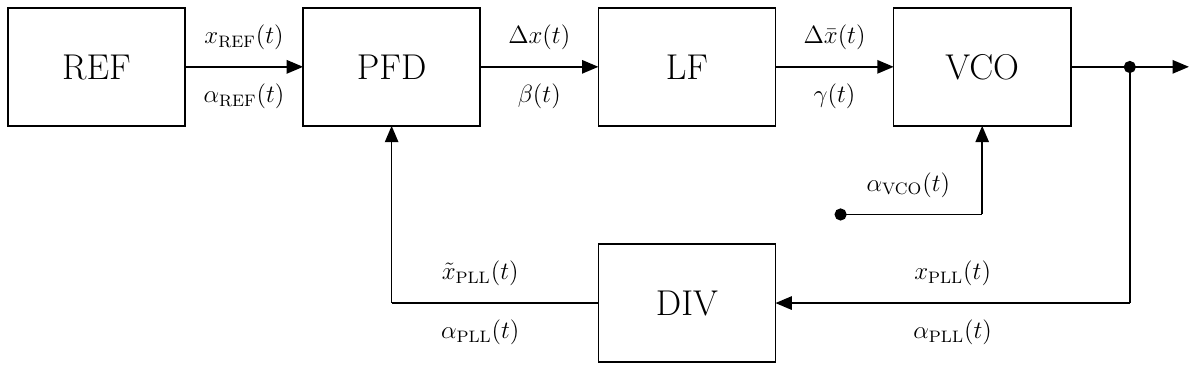}
    \caption{\ac{PLL} circuit diagram, REF reference signal oscillator, PFD phase frequency detector, LF loop filter, VCO voltage controlled oscillator, DIV frequency divider}
    \label{fig:PLL_circuit}
\end{figure}

In this section the description of a free running oscillator is extended to a \ac{PLL} circuit. 
In this part, a system model for the phase noise of a \ac{PLL} is established.
A \ac{PLL} is typically used in the RF chain of a \ac{SDR} device to synthesize the desired carrier frequency (cf. Fig. \ref{fig:usrp2953_block_diagram}).
The \ac{PLL} circuit is a control loop that combines the dynamic adjustability of a \ac{VCO}'s output frequency with the favorable phase noise characteristics of a reference oscillator. 

\subsection{Phase-Locked Loop Model - Components and Basics}

Fig. \ref{fig:PLL_circuit} shows the \ac{PLL} circuit with all its components.
In the figure, $x_\text{REF},\Delta x, \Delta \Bar{x}, x_\text{PLL}, \Tilde{x}_\text{PLL}$ represent time dependent voltage signals.
The terms $\alpha_\text{REF}, \beta, \gamma, \alpha_\text{VCO}, \alpha_\text{PLL}$ refer to stochastic time shift processes affecting the voltage signals.
The output signal of the \ac{PLL} is given as 
\begin{align}
    x_\text{PLL}(t) &= \cos\left[ 2\pi f_\text{PLL} (t + \alpha_\text{PLL}(t))\right] \nonumber\\
               &= \cos\left[ 2\pi f_\text{PLL}t + \theta_\text{PLL}(t)\right].
\end{align}
The signal after the frequency divider is
\begin{align}
    \Tilde{x}_\text{PLL}(t) &= \cos\left[ 2\pi \underbrace{\frac{f_\text{PLL}}{M}}_{=f_\text{REF}} (t + \alpha_\text{PLL}(t))\right]\nonumber\\
                       &= \cos\left[ 2\pi f_\text{PLL}t + \Tilde{\theta}_\text{PLL}(t)\right].
\end{align}
The \ac{PFD} compares the phases of its input signals 
\begin{align}
    \Delta\theta(t) &= \Tilde{\theta}_\text{PLL}(t) - \theta_\text{REF}(t) \nonumber\\
                 &= 2\pi f_\text{REF} \underbrace{(\alpha_\text{PLL}(t) - \alpha_\text{REF})}_{=\beta(t)}(t).
\end{align}
Its output signal $\Delta x (t)$ is a pulse-width modulated signal whose duty cycle depends on $\Delta\theta(t) \sim \beta(t)$ and is scaled with the phase detector gain $k_\text{PFD}$
\begin{align}
    \Delta x (t) = k_\text{PFD} \sum_{k=0}^{\infty} \text{rect} \left( \frac{t - 0.5 \beta (\tfrac{k}{f_\text{REF}})}{\beta (\tfrac{k}{f_\text{REF}})} \right) \ast \delta (t -\tfrac{k}{f_\text{REF}}).
\end{align}
The idea of a \ac{LF} is to balance the effect of the previous differentiation step in the \ac{PFD} to improve the stability of the control loop.
Therefore, a \ac{LF} is typically a filter that can be a \ac{LPF} of first order, second order or a charge pump that sums up the error signal $\Delta x (t)$.
Its output is a \ac{LP}-filtered error signal $\Delta \Bar{x} (t) = h_\text{LF}(t) \ast \Delta x(t)$ with corresponding time shift process $\gamma (t)$.
The \ac{VCO} input is the filtered error signal $\Delta \Bar{x} (t + \gamma (t))$ as the control voltage, affected by a corresponding filtered stochastic time shift.
Furthermore, the \ac{VCO} has its own phase noise process, represented by the stochastic time shift $\alpha_\text{VCO}(t)$.\\
\begin{figure*}[t]
{
    \centering
    \includegraphics[width=0.8\textwidth]{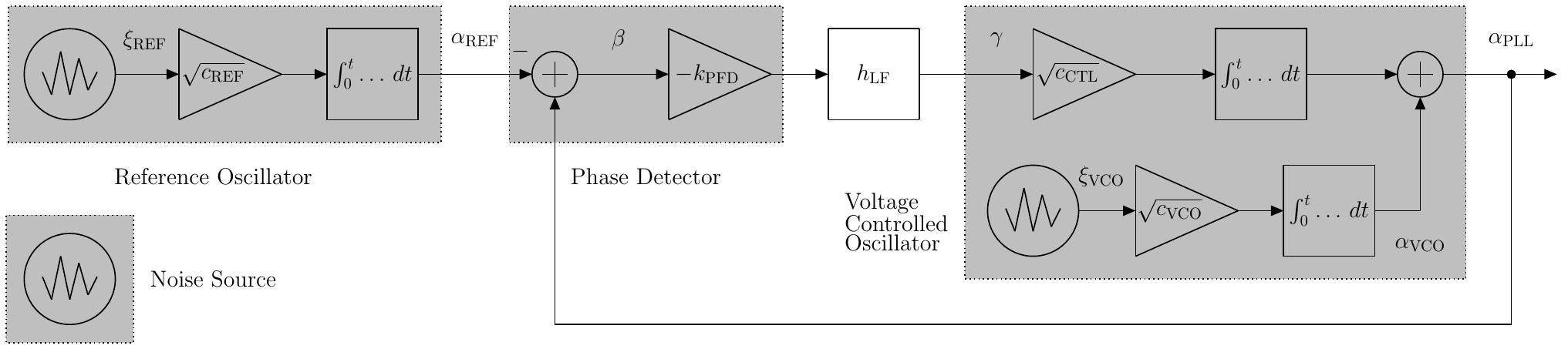}
    \caption{Noise equivalent block diagram for \ac{PLL} \cite{Coll202501}, considering white noise sources effecting the control voltage of reference oscillator and \ac{VCO}}
    \label{fig:noise_equivalent_circuit_PLL}
    }
\end{figure*}

The assumptions for the considered \ac{PLL} model are \cite{mehrotra1031966}:
\begin{itemize}
    \item[1.] The \ac{PLL} is locked, i.e. $f_\text{REF} \approx f_\text{PLL}/M$, with frequency of reference signal $f_\text{REF}$, \ac{PLL} output frequency $f_\text{PLL}$ and divider ratio $M$.
    \item[2.] The input of the \ac{PLL} is a reference signal source which is modeled as a free running oscillator)
    \begin{align}
        x_\text{REF}(t) &= \cos\left[ 2\pi f_\text{REF} (t + \alpha_\text{REF}(t))\right]\nonumber\\
                   &= \cos\left[ 2\pi f_\text{REF} t + \theta_\text{REF}(t))\right].
    \end{align}
    \item[3.] All the noise sources in the \ac{PLL} circuit are zero mean white Gaussian and are independent of each other. 
\end{itemize}

\subsection{Phase Noise in Phase-Locked Loops}

Fig. \ref{fig:noise_equivalent_circuit_PLL} shows the noise equivalent block diagram for the \ac{PLL}.
It is assumed that other components (divider, loop filter, etc.) besides the oscillators are noise free.
In the simple case, where no loop filter is present and we consider two noise sources ($p=1, q=2$), the system of differential equations can be written as (c.f. \eqref{eq:diff_pll})
\begin{align}
    &\frac{d}{dt}\underbrace{\beta(t)}_{\mathbf{y}} = -\underbrace{\sqrt{c_\text{CTL}}k_\text{PFD}}_{\mathbf{A}} \underbrace{\beta(t)}_{\mathbf{y}}\nonumber\\
    &+ 
    \underbrace{
    \begin{bmatrix}
    \sqrt{c_\text{VCO}} & -\sqrt{c_\text{REF}}
    \end{bmatrix}}_{\mathbf{B}}
    \underbrace{
    \begin{bmatrix}
        \xi_\text{VCO}(t)\\
        \xi_\text{REF}(t)
    \end{bmatrix}}_{\mathbf{\xi}}.
    \label{eq:diff_pll_rank_one}
\end{align}
The characterization of a generic \ac{PLL} and derivation of the system of differential equations representing the \ac{PLL} dynamics is shown in Appendix~\ref{appendix:pn_in_pll}.
In the considered case of a first order \ac{PLL}, with ${\sqrt{c_\text{CTL}}k_\text{PFD}=\omega_\text{PLL}=2\pi f_\text{PLL}}$, the solution to \eqref{eq:diff_pll_rank_one} is
\begin{align}
    \beta(t) &= \int_{0}^{t} e^{-2 \pi f_\text{PLL} (t - t')} \left( \sqrt{c_\text{VCO}} \xi_\text{VCO}(t) - \sqrt{c_\text{REF}} \xi_\text{REF}(t) \right) dt.
    \label{eq:pll_sde_beta}
\end{align}

\subsection{Discrete Time Domain Model}
The model described here represents a \ac{PLL} in the discrete time.
The time index is represented by variables $i,n$ and the sample interval is denoted by $\Delta t$.
In discrete time, the time shift process at the \ac{PLL} output can be expressed as (cf. \eqref{eq:diff_pll_rank_one} by  substituting $\alpha[i] = \sqrt{c} \int_0^i \xi(t)dt$), such that
\begin{align}
    \alpha_\text{PLL}[n] = 
    \begin{cases}
      0, & n = 0 \\
       \sum_{i=0}^{n-1} \left( \alpha_\text{PLL}[i] - \alpha_\text{REF}[i] \right)& \\
      \cdot \left( -2\pi f_\text{PLL}\Delta t \right) + \alpha_\text{VCO}[n-1], & n > 0
    \end{cases}.
    \label{eq:pn_pll}
\end{align}
Note that this process becomes stationary when $\alpha_\text{REF}$ becomes constant.
Identical to the case of a free running oscillator, the corresponding phase shift process at \ac{PLL} output becomes $\phi_\text{PLL}[n] = 2 \pi f_0 \alpha_\text{PLL}[n]$.

\subsection{Autocorrelation Function of Phase-Locked Loop Output Process}

Similar to the approach for describing a free running \ac{VCO} previously, the following step is to obtain the \ac{ACF} for the process at the \ac{PLL} output.
With \eqref{eq:pll_output_process} and \eqref{eq:acf_alpha} the \ac{ACF} can be written as
\begin{align}
    R_{\alpha,\alpha}(t, t + \tau) &= \mathbb{E} \left[ \alpha_\text{PLL}(t) \alpha_\text{PLL}(t+\tau) \right]\nonumber\\
                                   &= c_\text{REF} \text{min}(t, t+\tau) + R_{\alpha,\beta}(t, t+\tau)\nonumber\\
    &+ R_{\beta,\alpha}(t, t+\tau) + R_{\beta,\beta}(t, t+\tau).
\label{eq:acf_pll_out_exp}
\end{align}
Further derivation steps are given in the Appendix~\ref{appendix:acf_pll_out}.
The \ac{ACF} of $\alpha_\text{PLL}$ at the \ac{PLL} output is
\begin{align}
    &R_{\alpha,\alpha} (t, t + \tau) = \frac{c_\text{VCO} + c_\text{REF}}{4 \pi f_\text{PLL}} \left[ e^{-2 \pi f_\text{PLL} |\tau|} - e^{-2 \pi f_\text{PLL}(2t + \tau)} \right] \nonumber\\
    &- \frac{c_\text{REF}}{2 \pi f_\text{PLL}} \left[  e^{-2 \pi f_\text{PLL} |\tau|} - e^{-2 \pi f_\text{PLL} (t + \tau)} + 1  - e^{-2 \pi f_\text{PLL} t }\right]\nonumber\\
    &+c_\text{REF} t.\label{eq:acf_pll}
\end{align}
\subsection{Variance of Phase-Locked Loop Output Process}
\begin{figure}[tb]
    \centering
    \includegraphics[width=0.8\linewidth]{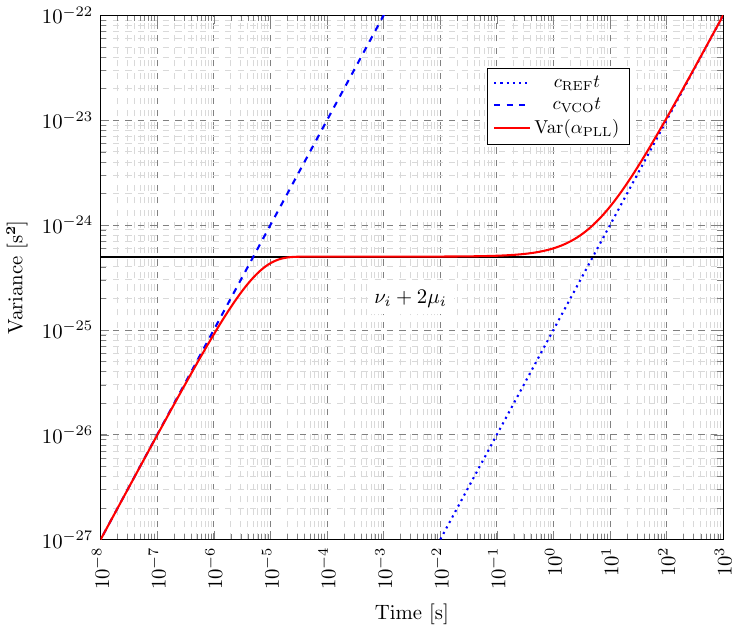}
    \caption{Variances for \ac{PLL} and free running oscillators REF and \ac{VCO} for $c_\text{REF} = \SI{e-16}{s}, c_\text{VCO} = \SI{e-14}{s}, 2 \pi f_\text{PLL} = \SI{e5}{rad \cdot s^{-1}}$}
    \label{fig:phase_noise_pll_var}
\end{figure}

In order to illustrate the key characteristic of the \ac{ACF}, in this subsection the Variance of the \ac{PLL} time shift output process is plotted in Fig. \ref{fig:phase_noise_pll_var}.
The variance of the output process $\alpha_\text{PLL}$ is found for $\tau=0$ of its \ac{ACF} \eqref{eq:acf_pll}
\begin{align}
    &\text{Var}(\alpha_\text{PLL}(t)) = R_{\alpha,\alpha} (t, \tau=0) .
\end{align}
In the case of a first order \ac{PLL} (no \ac{LF}) the expression becomes 
\begin{align}
    \text{Var}(\alpha_\text{PLL}) &= \frac{c_\text{VCO} + c_\text{REF}}{4 \pi f_\text{PLL}} \left[ 1 - e^{-4 \pi f_\text{PLL} t}\right] \nonumber\\
    &- 2 \frac{c_\text{REF}}{2 \pi f_\text{PLL}} \left[ 1 - e^{-2 \pi f_\text{PLL} t}\right] + c_\text{REF} t \nonumber\\
    &\approx \frac{c_\text{VCO} - 3 c_\text{REF}}{4 \pi f_\text{PLL}} \left[ 1 - e^{-4 \pi f_\text{PLL} t} \right] + c_\text{REF} t.
    \label{eq:var_pll}
\end{align}
Fig. \ref{fig:phase_noise_pll_var} illustrates the behaviour of the variance given by \eqref{eq:var_pll}.
For smaller times, the \ac{PLL} variance follows that of a free running oscillator \ac{VCO}.
Between $\SI{e-5}{s}$ to $\SI{e-1}{s}$ the variance remains constant at ${\nu_i + 2 \mu_i = \frac{c_\text{VCO} - 3 c_\text{REF}}{4 \pi f_\text{PLL}}}$.
Note that this behaviour is explained by \ac{OU} process $\beta(t)$ having a constant variance.
At larger times (${> \SI{e2}{s}}$) the contribution from the $c_\text{REF} t$ term in \eqref{eq:var_pll} becomes significant and the variance of the \ac{PLL} follows that of the \ac{REF} oscillator.
\subsection{Power Spectrum Density of Phase-Locked Loop}
For deriving the phase noise spectrum of the \ac{PLL}, a similar approach to previous derivation for the \ac{VCO} will be taken.
The full derivation is provided in Appendix~\ref{appendix:psd_pll}.
Similar to \eqref{eq:fourier_series_ideal_osc}, a Fourier series expression for the ideal oscillator signal is used.
The \ac{ACF} of the \ac{PLL} output signal $x_\text{PLL}(t) = x_0(t + \alpha_\text{PLL}(t))$ around the first harmonic is
\begin{align}
    R_{x,x}(t, t + \tau) = \mathbb{E} [x_0(t + \alpha_\text{PLL}(t)) x_0^* (t + \tau + \alpha_\text{PLL}(t + \tau)) ].
\end{align}
Then, the asymptotic \ac{ACF} of the $\alpha_\text{PLL}$ process can be written as
\begin{align}
    &\lim_{t\to\infty} R_{x,x}(t, t + \tau) =  \sum_{i=\pm 1} \left| \underline{X}_i \right|^2 e^{ji 2 \pi f_0 \tau}\nonumber\\
    &\cdot e^{- (2 \pi f_0)^2 \left[ 
    \frac{c_\text{VCO} - c_\text{REF}}{4 \pi f_\text{PLL}} \left[ 1 - e^{-2 \pi f_\text{PLL} |\tau|} \right] + 0.5 c_\text{REF} |\tau| \right]}.
    \label{eq:pll_acf_asym}
\end{align}
The \ac{PSD} of the \ac{PLL} output is 
\begin{align}
    &S_{x,x}(f) = \mathfrak{F}_\tau \Bigl\{ R_{x,x} (\tau) \Bigr\}\nonumber\\
    &= \sum_{i=\pm 1} \left| \underline{X}_i \right|^2 \sum_{n=0}^{\infty} \frac{1}{n!} e^{- \frac{\pi f_\text{0}^2}{f_\text{PLL}} \left( c_\text{VCO} - c_\text{REF} \right)}\nonumber\\
    &\cdot {\left( \frac{\pi f_\text{0}^2}{f_\text{PLL}} \left( c_\text{VCO} - c_\text{REF} \right) \right)}^{n} \nonumber\\
    &\cdot \frac{4 \pi \left( \pi f_\text{0}^2 c_\text{REF} + n f_\text{PLL} \right)}{\left(2 \pi \left( \pi f_\text{0}^2 c_\text{REF} + n f_\text{PLL}  \right) \right)^2 + (f - i f_\text{0})^2}. \label{eq:pll_pdf_out}
\end{align}
Comparing the phase noise spectrum for the output of a noisy \ac{PLL} to that of a free running oscillator \eqref{eq:phase_noise_psd_omega}:
\begin{itemize}
    \item The expression of the spectrum consists a Dirac-pulse $\delta(f)$ for the ideal oscillator spread by phase noise.
    \item The spectrum consists of a (weighted) sum of Lorentzian spectrum's, shifted by frequency $f_\text{0}$ of the \ac{VCO} in the \ac{PLL}
\end{itemize}
\begin{figure}[tb]
    \centering
    \includegraphics[width=0.8\linewidth]{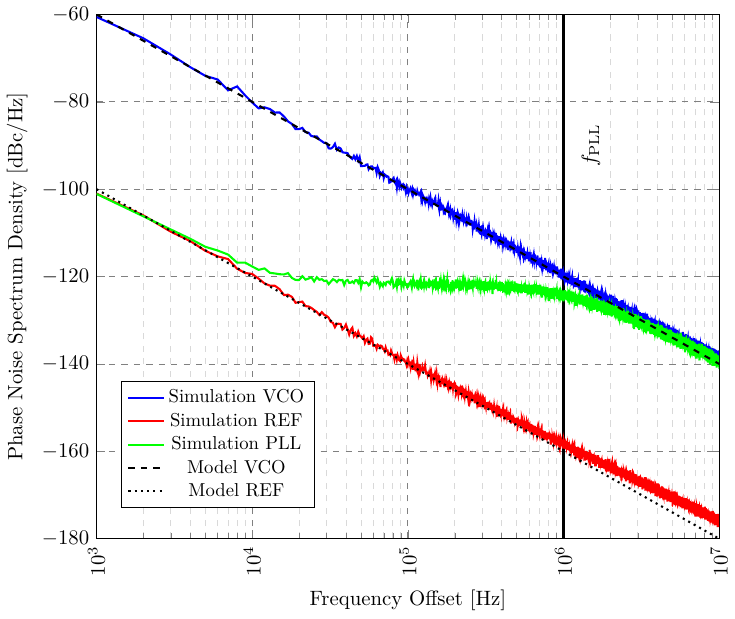}
    \caption{Phase noise of \ac{PLL}, free running \ac{VCO} and \ac{REF}, parameters $c_\text{REF} = \SI{e-16}{s}, c_\text{VCO} = \SI{e-14}{s}, f_\text{PLL}=k_\text{PFD} \sqrt{c_\text{CTL}}/2\pi = \SI{e6}{Hz}, f_s= \SI{e8}{Hz}$ and $N=\SI{100}{k}$ samples} 
    \label{fig:phase_noise_pll}
\end{figure}
In Fig.~\ref{fig:phase_noise_pll}, the phase noise spectrum of the \ac{PLL} circuit is shown.
Note that for low frequencies, the \ac{PLL} follows the behavior of the free running \ac{REF} oscillator.
The \ac{PLL} bandwidth $f_\text{PLL}$ characterizes the frequency after which it follows the behavior of a free running \ac{VCO} asymptotically.
During the transition interval up to the \ac{PLL} bandwidth, the spectrum is nearly constant.
For the model of \ac{VCO} and \ac{REF} the general derivation for a free running \ac{VCO}  in \eqref{eq:phase_noise_psd_f_offs} is used.
\subsection{Simplified Expression for Phase Noise Spectrum of Phase-Locked Loops}
\begin{figure}[tb]
    \centering
    \includegraphics[width=0.8\linewidth]{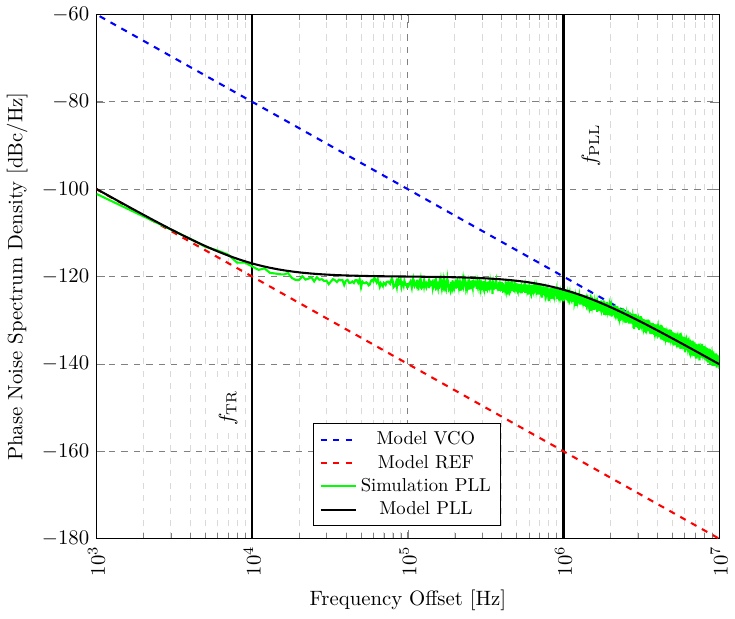}
    \caption{Phase noise of \ac{PLL}, comparing simplified phase noise model with simulation, identical parameters to Fig. \ref{fig:phase_noise_pll}, $f_\text{TR}=\SI{e4}{Hz}$} 
    \label{fig:phase_noise_pll_sim}
\end{figure}
The next step is simplifying  the expression \eqref{eq:pll_pdf_out} for the \ac{PLL} output \ac{PSD}.
The sum of Lorentzian spectrums shall be reduced to the minimum number of necessary components.
From Fig.~\ref{fig:phase_noise_pll_sim}, it can be seen that for low offset frequencies, the PLL follows the characteristic of the \ac{REF} oscillator.
The \ac{REF} oscillators phase noise is described by (cf. \eqref{eq:phase_noise_psd_f_offs}, units ommited)
\begin{align}
    \mathfrak{L}_\text{REF}(f_\text{offs}) &= 10 \log_{10} \left( \frac{f_{0}^2 c_\text{REF}}{f_\text{offs}^2 + \pi^2 f_{0}^4 c_\text{REF}^2} \right)\nonumber\\
    &= 10 \log_{10} \left( \frac{1}{\pi^2 f_{0}^2 c_\text{REF}} \frac{1}{1 + \frac{f_\text{offs}^2}{f_\text{3dB,REF}^2}} \right).
    \label{eq:l_ref}
\end{align}
During the transition interval, starting at frequency $f_\text{TR}$ the spectrum remains constant until the \ac{PLL} bandwidth $f_\text{PLL}$.
For higher frequencies, \ac{PLL} follows the characteristic of the \ac{VCO}, given by (cf. \eqref{eq:phase_noise_psd_f_offs})
\begin{align}
    \mathfrak{L}_\text{VCO}(f_\text{offs}) &= 10 \log_{10} \left( \frac{1}{\pi^2 f_{0}^2 c_\text{VCO}} \frac{1}{1 + \frac{f_\text{offs}^2}{f_\text{3dB,VCO}^2}} \right).
    \label{eq:l_vco}
\end{align}
The unknown frequency $f_\text{TR}$ can be obtained by first calculating $\mathfrak{L}_\text{VCO}(f_\text{offs}=f_\text{PLL})$.
Then, the equation $\mathfrak{L}_\text{VCO}(f_\text{offs}=f_\text{PLL}) = \mathfrak{L}_\text{REF}(f_\text{offs}=f_\text{TR})$ can be solved for $f_\text{TR}$.
For the choice of parameters given in caption Fig. \ref{fig:phase_noise_pll} we obtain $f_\text{TR} = \SI{e4}{Hz}$.
To have a more general expression the exponents in each component are expressed as a variable $k_\text{VCO}, k_\text{REF}$ for the slope of the \ac{LP} filters representing the \ac{VCO} and \ac{REF}.
In our derivation, assuming white noise sources, the variables are $k_\text{VCO} = k_\text{REF} = 2$, which leads to a slope of $\SI{-20}{\frac{dBc}{Hz \cdot decade}}$.
However, this might not reflect the characteristic of a specific \ac{VCO} or \ac{REF} appropriately, which necessitates a generic expression.
Finally the simplified expression for the phase noise spectrum can be expressed as
\begin{align}
    &\mathfrak{L}_\text{PLL}(f_\text{offs}) = 10 \log_{10} \left[ \frac{1}{\pi f_\text{3dB,REF}}\right. \nonumber\\
    & \cdot \frac{1 + \left( \frac{f_\text{offs}}{f_\text{TR}}\right)^{k_\text{REF}}}{1 + \left( \frac{f_\text{offs}}{f_\text{3dB,REF}} \right)^{k_\text{REF}}} \cdot \left. \frac{1}{1 + \left(\frac{f_\text{offs}}{f_\text{PLL}}\right)^{k_\text{VCO}} } \right].
    \label{eq:pn_model_simple}
\end{align}
This \ac{PLL} model is illustrated in Fig.~\ref{fig:phase_noise_pll_sim} by the black plot. 

	\section{Measurement}\label{sec:measurement}
The aim in the following parts is to provide a framework for fitting the previously outlined system model to specific hardware.
This is done to create a meaningful connection between theory and practical reality.
To achieve this, measurements of the phase noise spectrum of common \ac{SDR} devices are recorded.
The SDR investigated in this work are variants of the commonly in mobile communication research and development used NI USRP X310.
\subsection{Measurement Setup}
\begin{figure}[bt]
    \centering
    \includegraphics[width=\linewidth]{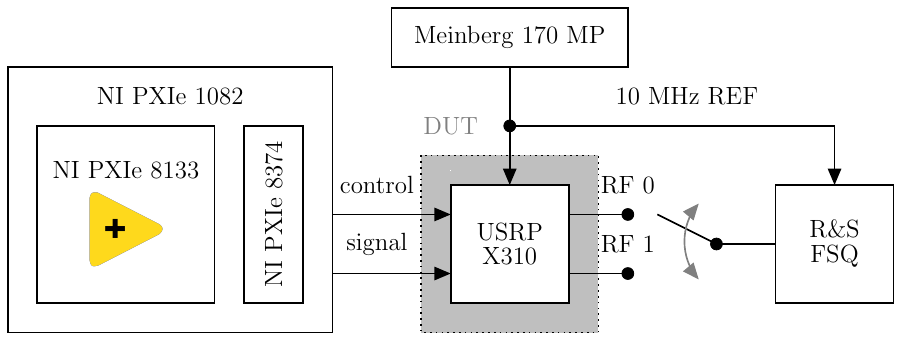}
    \caption{Setup for measuring the phase noise spectrum of \ac{SDR} \ac{USRP} X310 with reference clock Meinberg 170 MP}
    \label{fig:pn_meas_setup}
\end{figure}
Fig.~\ref{fig:pn_meas_setup} shows the phase noise measurement setup.
The LabVIEW program running on the NI PXIe 8133 controller generates the digital signal and provides control over signal parameters.
This signal is streamed to the \ac{SDR} device over the NI PXIe 8374 interface.
The streaming interface and controller are mounted in a NI PXIe 1082 chassis.
The LabVIEW controller provides control over signal parameters such as bandwidth, carrier frequency and power/gain.
For the measurements, a constant value baseband signal is generated to allow observation of the signal synthesized by the \ac{PLL}.
Both the \ac{USRP} and the spectrum analyzer R\&S FSQ used to perform the phase noise spectrum measurement are connected to the same 10 MHz \ac{REF} from a Meinberg 170 MP GPS receiver.
The common reference oscillator is essential to ensure frequency synchronicity between \ac{DUT} and R\&S FSQ.
The \ac{USRP} \ac{RF} port, for which the signal is generated, is directly connected to the R\&S FSQ via coaxial cable.\\
\subsection{Measurement Setup Configuration}
\begin{table}[ht]
    \centering
\begin{center}
        \resizebox{0.6\linewidth}{!}{
\begin{tabular}{| l | c |}
    \hline
    Property & Value\\
    \hline
    Carrier freq. (MHz) & \{500, 2000, 5000\}\\
    Bandwidth (MHz) & 10\\
    Power (dBm) & 0\\
    Waveform & constant value\\
    Port & \{RF0, RF1\} TX\\
    Reference (MHz) & 10\\
    \hline
\end{tabular}
}
\end{center}
\caption{\ac{USRP} configuration} 
\label{table:usrp_config}
\end{table}
The key parameters of the \ac{USRP} configuration are outlined in Table \ref{table:usrp_config}.
The carrier frequencies are selected from a broad range to capture the performance characteristics across different operating conditions.
Since the digital baseband signal is set as a constant value to obtain the oscillator signal at the \ac{RF} ports, the bandwidth plays a minor role and is set arbitrarily to $\SI{10}{MHz}$.
The phase noise spectrum measurement is performed for both \ac{RF} chains, due to each \ac{USRP} being equipped with two \ac{RF} frontend daughterboards.


%
\subsection{Phase Noise Spectrum Measurements}
In this subsection, the phase noise measurements for various \ac{USRP} models in different configurations is detailed.
The measured dataset \cite{data_bpmg_pc85_24} of phase noise spectrums contains data for the following \ac{USRP} models: 
\begin{itemize}
    \item USRP 2944R (\SI{160}{MHz}, UBX daughterboard) 
    \item USRP 2953R (\SI{120}{MHz}, CBX daughterboard) 
    \item USRP 2953R (\SI{40}{MHz},\hspace{0.5em} CBX daughterboard) 
    \item USRP 2954R (\SI{160}{MHz}, UBX daughterboard) 
\end{itemize}
In separate measurements, a \SI{30}{dB} attenuator is placed in front of the FSQ's input, to explore the effect on measured phase noise.
\begin{figure*}[tb]
    \begin{minipage}{0.32\textwidth}
        \centering
        \includegraphics[width=\linewidth]{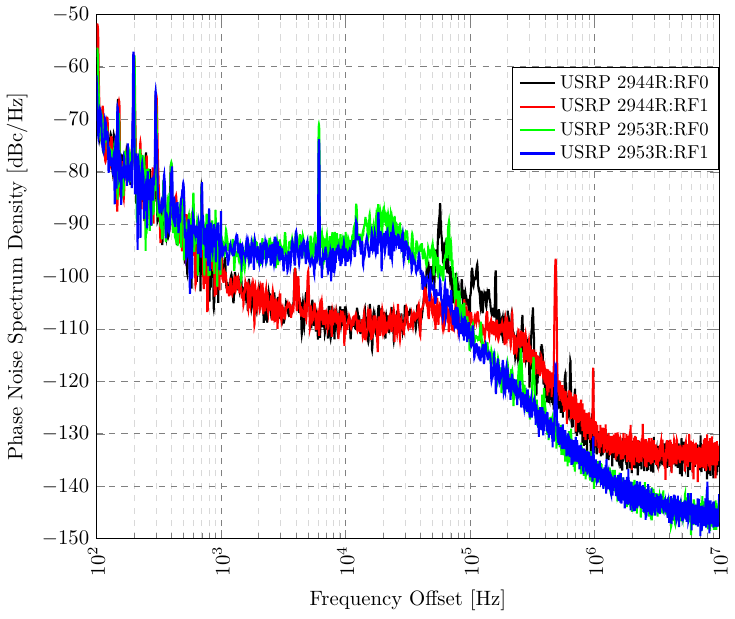}
        \caption{Measured phase noise spectrum \cite{data_bpmg_pc85_24} for USRP with different daughterboards - 2944R (UBX) and 2953R (CBX) at $f_\text{c} = \SI{2}{GHz}$}
        \label{fig:pn_meas_1}
    \end{minipage}
    \begin{minipage}{0.32\textwidth}
        \centering
        \includegraphics[width=\linewidth]{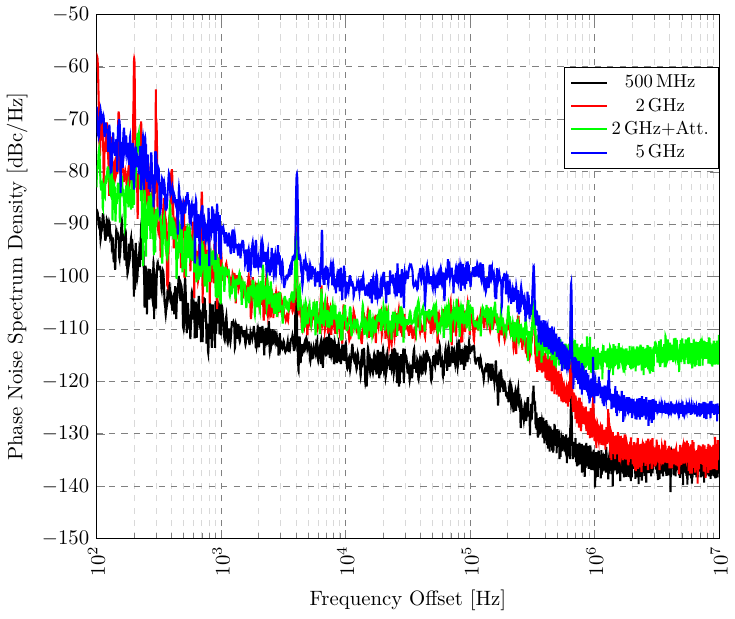}
        \caption{Measured phase noise spectrum \cite{data_bpmg_pc85_24} for \ac{USRP} 2944R (UBX) at different frequencies}
        \label{fig:pn_meas_2}
    \end{minipage}
    \begin{minipage}{0.32\textwidth}
        \centering
        \includegraphics[width=\linewidth]{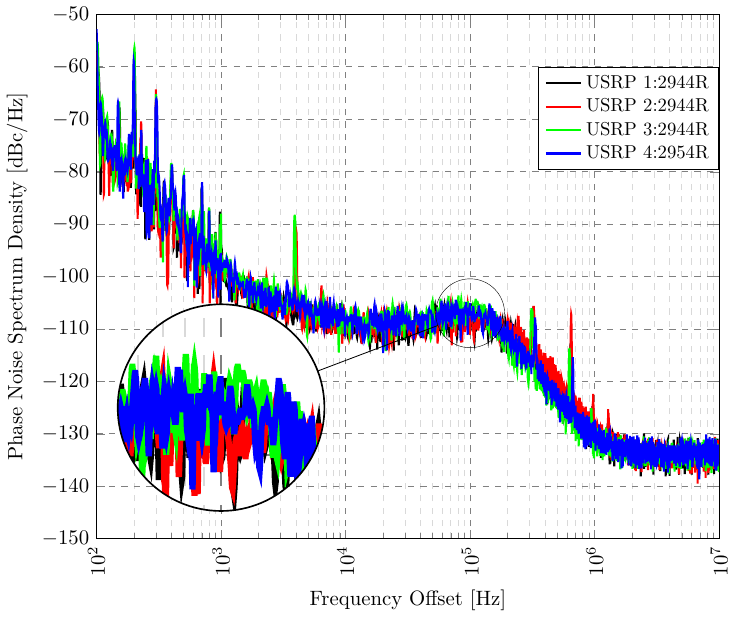}
        \caption{Measured phase noise spectrum \cite{data_bpmg_pc85_24} of different \ac{USRP} with identical daughterboards at $f_\text{c} = \SI{2}{GHz}$}
        \label{fig:pn_meas_3}
    \end{minipage}
\end{figure*}

In Fig.~\ref{fig:pn_meas_1}, \acp{USRP} with UBX and CBX daughterboards are compared.
It can be seen that while the \ac{PLL} bandwidth $f_\text{PLL}$ and $f_\text{TR}$ are slightly different, however, the overall characteristic of the phase noise spectrum remains similar.
This is expected given that the phase noise characteristics of their frequency synthesizing circuits (Maxim Integrated MAX2870 for CBX and MAX2871 for UBX) are identical\cite{MAX2870, MAX2871}.
The observed difference in the phase noise spectrum likely originate from different wiring and circuitry connecting the \ac{PLL} and other \ac{RF} components on the daughter- and motherboards, more specifically the \ac{LF}. 

In Fig. \ref{fig:pn_meas_2} the effect of different carrier frequencies on phase noise \ac{PSD} is explored.
The observable increase of the \ac{PSD} results from the phase noise scaling multiplicatively with oscillator frequency \eqref{eq:pn_model_simple}. 
In the green plot, a $\SI{30}{dB}$ attenuator is introduced to assess its impact on phase noise measurements. 
It can be seen that this effectively raises the noise floor from $\SI{-135}{dBc/Hz}$ to $\SI{-115}{dBc/Hz}$ in the measurement.
This results from the attenuator reducing the signal power fed to the FSQ.
Since the FSQ has an inherent fixed noise floor level given by its hardware properties, the effect is a raised observed noise floor in the measurement.
However, the overall phase noise characteristic of the \ac{USRP} output signal is not affected by the attenuator.
This is due to the \ac{PSD} representing the power density relative to the carrier power.
The relation is maintained as long as the absolute signal power does not fall below the detection limit of the FSQ, at which point the \ac{PSD} gradually submerges in the noise floor.

In Fig. \ref{fig:pn_meas_3} the phase noise spectrum of different \ac{USRP} with identical daughterboards UBX is shown.
It can be observed that the characteristics of the \ac{PSD}'s are practically identical, suggesting that phase noise behaviour is consistent among different \ac{USRP} with the same daughterboard type. 
This insight is important when phase coherent operation of multiple \ac{USRP}'s is desired.
Sufficiently similar phase noise characteristics are important to simplify phase synchronization efforts.
This is due to typical synchronization methods, such as suppression of phase noise with Kalman filtering relying on estimating the statistical moments of the phase noise processes \cite{Petrovic2003PhaseNS}.
When the spectral characteristics are distinct, these statistical moments can be significantly different, hindering phase synchronization efforts.

\subsection{Discussion}

This subsection is dedicated to discussing the differences and their causes between the model \eqref{eq:pn_model_simple}, highlighted in Fig. \ref{fig:phase_noise_pll_sim} and the measured \ac{PSD}'s in Fig. \ref{fig:pn_meas_1}, \ref{fig:pn_meas_2}, \ref{fig:pn_meas_3}.
While the overall spectrum shape in measurements is similar, specific differences can be noted.
Most notable is a difference in the slope of the spectrum ($\SI{-30}{\frac{dBc}{Hz \cdot decade}}$ in measurements and $\SI{-20}{\frac{dBc}{Hz \cdot decade}}$ in the model).
The slope observed in the measurements is identical to the datasheets \cite[p. 5]{MAX2870} \cite[p. 6]{MAX2871} in \textit{"VCO open loop phase noise"} for the \ac{PLL} circuits, which supports the credibility of the measurement. 
This difference results from other influences, besides white noise sources affecting the control voltage, affecting the phase noise characteristic of \ac{VCO} and \ac{REF} oscillators.
However, using the generic simplified model \eqref{eq:pn_model_simple} allows to fit the parameters $k_\text{VCO}, k_\text{REF}$ to match the specific characteristics of the oscillators.
The transition interval between \ac{REF} and \ac{VCO} dominant parts in the measured \ac{PSD} is not completely flat as in the model.
This difference occurs since the derivation assumed absence of a \ac{LP} filter for simplicity.
The datasheet figures \cite[p. 5]{MAX2870} \cite[p. 6]{MAX2871} in \textit{"closed loop phase noise"} exhibit identical characteristics to our measurements.
Another difference is that the measurement contains spurs, while the model does not account for this effect.
The spurs visible in a \ac{PLL}'s \ac{PSD} originate from various sources such as non-linearities in circuits, unwanted mixing products and over-driving \ac{RF} components \cite{xie2017fixedspur}.
Furthermore, the measured \ac{PSD} shows less smoothness in its low-frequency components compared to the model, primarily due to the limited measurement time-frame.
The total sweep time of $\SI{25}{s}$ is evidently not sufficient so that low frequency components are adequately represented in the measured spectrum. 
A practical solution to address this limitation is to record multiple measurements and average the measured \acp{PSD} in post-processing.
	\section{Parameter Estimation}\label{sec:parameter}
In this section, the measured \acp{PSD} are used to estimate the parameters of the simplified model \eqref{eq:pn_model_simple}.
The primary interest is in the oscillator constant for \ac{VCO} and \ac{REF}, $c_\text{VCO},c_\text{REF}$ and the corresponding $\SI{3}{dB}$ cut-off frequencies, frequency at the start of the transition interval $f_\text{TR}$ and \ac{PLL} bandwidth $f_\text{PLL}$.
The established system model is extended to account for the noise floor that was not considered in the derivation. 
Estimators for the system model parameters are presented and explicitly calculated for the recorded data-set.
Note that  the notations used in the system model \eqref{eq:pn_model_simple} are shortened here to $\mathfrak{L}_\text{PLL} = \mathfrak{L}$ and $f_\text{offs} = f$.
Furthermore, the accent $\hat{f}_{\text{TR}}$ is used to indicate parameter estimates.

\subsection{System Model Extension}
\begin{figure}[t]
    \centering
    \includegraphics[width=0.8\linewidth]{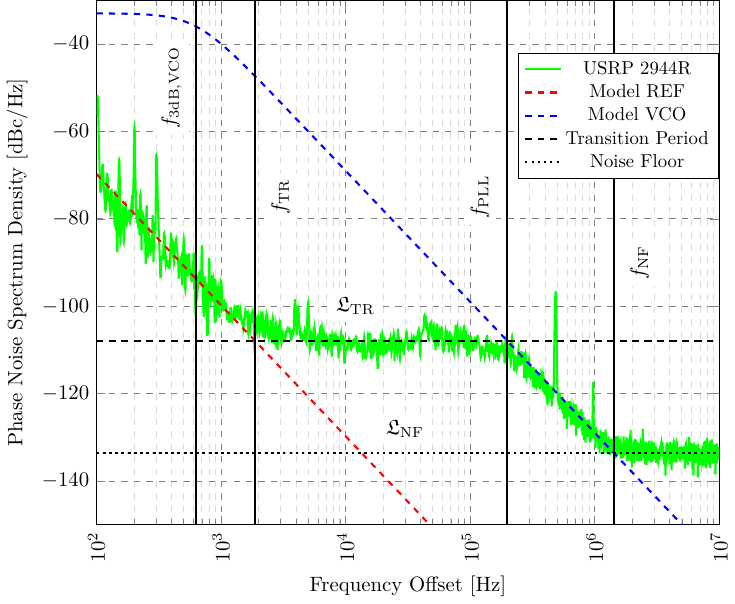}
    \caption{Phase noise spectrum of \ac{USRP} 2944R with parameter estimates and fitted models for \ac{REF} and \ac{VCO}}
    \label{fig:pn_param_est}
\end{figure}

%
Since the frequency response of all oscillators contains a noise floor, a noise floor is added to the system model with $\mathfrak{L}_\text{max} = - 10 \log_{10} (\pi f_\text{3dB,REF})$ (c.f. \ref{eq:l_max})
\begin{align}
    \label{eq:pn_model_simpl_noise_floor}
    \mathfrak{L}(f) &= \\ 
    \mathfrak{L}_\text{max} &+ 10 \log_{10} \left[ \frac{1 + {\left( \frac{f}{f_\text{TR}} \right)}^{k_\text{REF}}}{1 + \left( \frac{f^2}{f_\text{3dB,REF}^2} \right)^{k_\text{REF}}} \cdot \frac{1 + \left( \frac{f}{f_\text{NF}} \right)^{k_\text{VCO}}}{1 + \left( \frac{f^2}{f_\text{PLL}^2} \right)^{k_\text{VCO}}} \right].\nonumber
\end{align}
The noise floor of any oscillator can be observed at sufficiently large frequency offsets $f_\text{NF}$ relative to the oscillator frequency.
Note that the measured noise floor depends on the oscillator but also the sensitivity of the device used to measure the phase noise. 
The datasheet for the oscillator used on the UBX daughterboard \cite[p. 6]{MAX2871} and conducted measurements suggest that the slope of the phase noise spectrum is $\SI{-30}{\frac{dBc}{Hz \cdot decade}}$.
Therefore, the exponents in the system model are adjusted appropriately $k_\text{REF} = k_\text{VCO} = 3$.
While in \eqref{eq:pn_model_simpl_noise_floor} the $\SI{3}{dB}$ frequency of the \ac{VCO} $f_\text{3dB,VCO}$ is not explicitly included, it is implicitly contained in $f_\text{TR}$ which was introduced in \eqref{eq:pn_model_simple} to allow for more concise notation
\begin{align}
f_\text{TR} &= f_\text{3dB,REF} \sqrt{\frac{f_\text{3dB,VCO}^2 + f_\text{PLL}^2}{f_\text{3dB,VCO} \cdot f_\text{3dB,REF}} - 1}
\end{align}
in the case of $k_\text{VCO}=k_\text{REF}=2$ and in the general case
\begin{align}
f_\text{TR} &= \sqrt[k_\text{REF}]{\frac{f_\text{3dB,VCO}^{k_\text{VCO}} + f_\text{PLL}^{k_\text{VCO}}}{f_\text{3dB,VCO}}  f_\text{3dB,REF} - f_\text{3dB,REF}^{k_\text{REF}}}.
\end{align}
Note that these expressions are derived from \eqref{eq:pn_model_simpl_noise_floor} by  neglecting non significant components and solving for the desired intersection point frequency (cf. \figref{fig:pn_param_est}). 

\subsection{Analyzing PSD Characteristics}

\begin{figure}[t]
    \centering
    \includegraphics[width=0.8\linewidth]{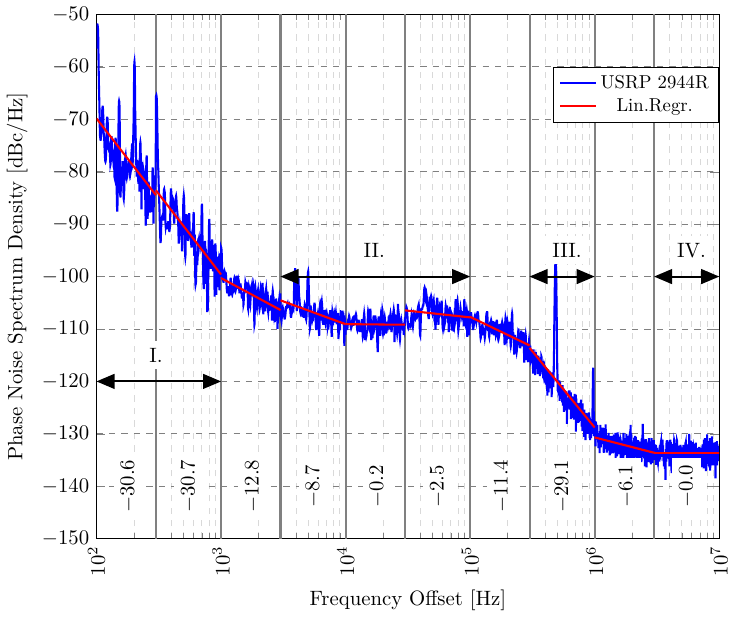}
    \caption{Phase noise spectrum of \ac{USRP} 2944R with UBX daughterboard in blue, piece-wise linear regression in red and corresponding estimates of slope $\hat{R}[1]$ in $\left[\SI{}{\frac{dBc}{Hz \cdot decade}}\right]$ given at the bottom of the plot}
    \label{fig:pn_lin_regr}
\end{figure}

Before the parameters of the system model can be estimated, the \ac{PSD} has to be split up in the following characteristic parts: 1) \ac{REF} dominant influence, 2) transition interval, 3) \ac{VCO} dominant influence, and 4) noise floor.
To achieve this, the slope of the \ac{PSD} is estimated using linear regression.
Fig.~\ref{fig:pn_lin_regr} shows a piece-wise linear regression of the phase noise spectrum given by
\begin{align}
    \hat{\bm{R}} = \left( \textbf{X}^{T} \textbf{X} \right)^{-1} \textbf{X}^{T} \textbf{Y}.  
\end{align}
The matrix
\begin{align}
    \textbf{X} = \begin{bmatrix}
        1 & \cdots & 1\\
        \log_{10} (f_{0}) & \cdots & \log_{10} (f_{n})
    \end{bmatrix}
\end{align}
contains the logarithmized bounds of the segment for which the regression is calculated, while $\textbf{Y} = [\mathfrak{L}(f_{0}), \cdots, \mathfrak{L}(f_{n})]$ represents the corresponding magnitude values of the phase noise spectrum.
Clearly distinguishable are the following sections:
\begin{itemize}
    \item[1.] $f < \SI{1}{kHz}$: \ac{PLL} follows \ac{REF}
    \item[2.] $ \SI{3}{kHz} ~\text{to} \leq f\leq\SI{100}{kHz}$: \ac{PLL} in transistion period
    \item[3.] $\SI{300}{kHz} ~\leq f\leq ~\SI{1}{MHz}$: \ac{PLL} follows \ac{VCO}
    \item[4.] $f > \SI{3}{MHz}$: noise floor
\end{itemize}
\subsection{Estimators for Phase Noise Model Parameters}
\label{sec:param_parameter_est}
This subsection is dedicated to estimating the parameters mentioned in the start of the section.
Explicit numerical values for parameter estimates are given in Appendix~\ref{appendix:pn_model_params}.
To estimate the oscillator constants, it is necessary to first estimate the cut-off frequencies for the \ac{LP} representing the phase noise characteristic of \ac{VCO} and \ac{REF}.
A general expression for the $\SI{3}{dB}$ cut-off frequency estimate can be derived from \eqref{eq:l_ref}, 
\begin{align}
    \hat{f}_\text{3dB} = 10^{ \frac{1}{k-1} \sum_{n=0}^{N} \log_{10} \left( 10^{\mathfrak{L}_{n}/10} \pi f_{n}^k \right) }.
    \label{eq:3db_cut_off}
\end{align}
Note that the relation $f_\text{offs} = f >> f_\text{3dB}$ is used for simplification in separating the variable.
Furthermore, $\mathfrak{L}_n$ is equivalent to $\mathfrak{L}(f_n)$ and $k \in \left\lbrace k_\text{REF}, k_\text{VCO} \right\rbrace$.
By summing over logarithmic values, the effect of spurs on the parameter estimate is reduced.
\\
The estimator for the $\SI{3}{dB}$ cut-off frequency \eqref{eq:3db_cut_off} is used on the intervals where the \ac{REF} and \ac{VCO} are dominant.
For the measurement displayed in Fig. \ref{fig:pn_lin_regr}, this yields the estimate for the $\SI{3}{dB}$ cut-off frequency of \ac{REF} $\hat{f}_\text{3dB,REF}$ and \ac{VCO} $\hat{f}_\text{3dB,VCO}$
As the oscillator frequency during the measurement is known, the oscillator constants can be derived by the following relation
\begin{align*}
    \hat{c} = \frac{f_\text{3dB}}{\pi f_0^2}.
\end{align*}
%

The next objective is to estimate the frequency where the transition interval starts $f_\text{TR}$, \ac{PLL} bandwidth $f_\text{PLL}$ and the frequency where the noise floor starts $f_\text{NF}$.
To achieve this, first the power level of the transition interval and the noise floor is estimated.
This power level estimate is obtained via the sample mean estimator
\begin{align}
    \hat{\mathfrak{L}} = \frac{1}{N} \sum_{n=0}^{N} \mathfrak{L}_{n}.
\end{align}
Using this estimator, the estimated power levels for transition interval $\hat{\mathfrak{L}}_\text{TR}$ and noise floor $\hat{\mathfrak{L}}_\text{NF}$ are obtained.
The desired frequencies can then be found at the intersection point of the power levels $\hat{\mathfrak{L}}$ and the \ac{LP} filters modeling the \ac{REF} \eqref{eq:l_ref} and \ac{VCO} \eqref{eq:l_vco}.
The equation is rearranged to isolate the offset frequency $f$.
Then the estimator for the (offset) frequency can be written as
\begin{align}
    \hat{f} = \hat{f}_\text{3dB} \sqrt[k]{\left(\frac{1}{10^{\hat{\mathfrak{L}}/10} \pi \hat{f}_\text{3dB}} - 1 \right)}.
\end{align}
By inserting $\hat{f}_\text{3dB} = \hat{f}_\text{3dB,REF}$ and $\mathfrak{L} = \hat{\mathfrak{L}}_\text{TR}$, the transition interval start frequency $\hat{f}_\text{TR}$ is found.
In similar fashion, the \ac{PLL} bandwidth $\hat{f}_\text{PLL}$ and noise floor start frequency $\hat{f}_\text{NF}$ can be calculated.
%
\begin{figure}[tb]
    \centering
    \includegraphics[width=0.8\linewidth]{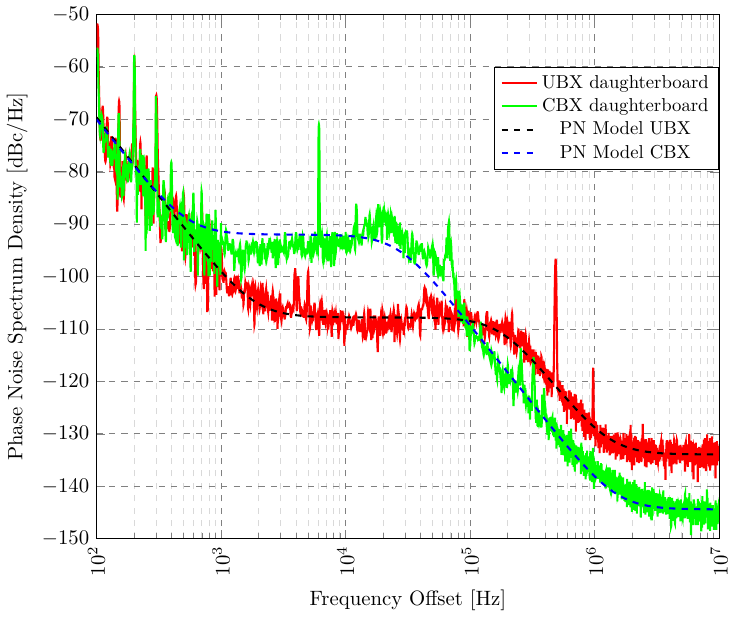}
    \caption{Phase noise spectrum of the system model using the derived parameter estimates for each daughterboard}
    \label{fig:pn_param_est_model}
\end{figure}

\subsection{Comparing Estimates for Different Devices}
The procedure for estimating the system model parameters is repeated for all \ac{USRP}'s in the measured dataset \cite{data_bpmg_pc85_24}.
Parameter estimates are calculated as average between all \ac{USRP} with identical daughterboard type.
Phase noise measurements for both \ac{RF} chains of the X310 models are considered.

As expected, the parameter estimates displayed in table \ref{table:parameter_estimates} relating to \ac{REF} are identical for both daughterboards.
This results from using the same reference (see Fig. \ref{fig:pn_meas_setup}).
The characteristics of the \ac{VCO} differ slightly for both daughterboards.
This can be explained by both daughterboards having different \ac{VCO} models as described in section \ref{sec:measurement}.
A significant difference occurs in start frequency of the transition interval and its magnitude.
Also the \ac{PLL} bandwidth for the UBX daughterboard is much larger ($\SI{177.3}{kHz} > \SI{26.6}{kHz}$) than that of the CBX daughterboard.
This bandwidth is important for characterizing the impulse response of the \ac{PLL}.
However, if the \ac{PLL} is assumed to be in lock, this parameter is of lower significance.
The measured noise floor shows a slight difference between daughterboard models.
It is important to note that this measurement does not necessarily reflect the noise floor that the frequency synthesizing circuit can achieve (cf. \cite{MAX2870}, \cite{MAX2871}).
The observed noise floor is influenced by the whole \ac{RF} circuitry and the measurement device used.

In \figref{fig:pn_param_est_model} the phase noise spectrum of the system model is compared to the measurements for both daughterboard models.
The parameters used are those given in \tabref{table:parameter_estimates}.
It is Graphically shown that there is the significant difference in the magnitude of the transition interval.


\section{Conclusion}\label{sec:conclusion}
In this paper, a parametric phase noise model for a generic \ac{PLL} is derived based on existing models from the literature.
This derivation is obtained by analyzing the \ac{PLL} dynamics in the time domain.
The phase noise characteristics of common \ac{SDR} devices, specifically the \ac{USRP} X310 from National Instruments, are measured.
The parameters of the derived phase noise model are fitted to the recorded measurements.

Key findings are the phase noise model parameters listed in  \tabref{table:parameter_estimates} for the two examined daughterboards of \ac{USRP} \texttt{X310} and estimators for phase noise model parameters from measured phase noise \ac{PSD} described in section \ref{sec:parameter}.
These estimates can be used with discrete phase noise models for free running oscillators and \ac{PLL}'s given in section \ref{sec:vco} and \ref{sec:pll} to generate phase noise processes in time domain with identical statistical properties to the measured devices.
Furthermore, the estimators given in section \ref{sec:parameter} can be used to estimate the phase noise model parameters of any \ac{PLL} or free running oscillator.


	\section*{Acknowledgments}
This work was supported by BMBF under the project KOMSENS-6G (16KISK124) and by
the Hexa-X-II project funded from SNS JU under Grant Agreement No 101095759.
    \appendices
    \section{Phase Noise in Phase-Locked Loops}
\label{appendix:pn_in_pll}

Previous works \cite{mehrotra1031966} have shown that phase noise analysis leads to a system of differential equations
\begin{align}
    \frac{d}{dt}\mathbf{y} = -\mathbf{A} \mathbf{y} + \mathbf{B} \mathbf{\xi}. \label{eq:diff_pll}
\end{align}
The vector $\mathbf{y}(t) = {[\beta(t), \gamma(t), \hdots]}^T$ represents state variables and is of length $p$. 
The order of \ac{PLL} $p$ is characterized by the order of \ac{LF} $p-1$.
The noise sources of \ac{PLL} are represented by the vector $\mathbf{\xi}(t)$ of length $q$.
With $q$ representing the number of independent noise sources.
The matrices $\mathbf{A, B}$ are of dimension $p \times p$ and $p \times q$ respectively.
Using (\ref{eq:wiener_process_diff}) we can rewrite (\ref{eq:diff_pll}) 
\begin{align}
    d\mathbf{y}(t) = -\mathbf{A} \mathbf{y}(t) dt + \mathbf{B} d\mathbf{W}(t),
\end{align}
where the vector $d\mathbf{W}(t)$ of length $q$ represents independent Wiener-processes.
This stochastic differential equation is known in the literature \cite{papoulis_pillai78965125, gardiner198589073165} as the Langevin equation.
Its solution is a \ac{OU} process \cite[p.109, 4.4.43]{gardiner198589073165}
\begin{align}
    \mathbf{y}(t) = \mathbf{y}(0) e^{-\mathbf{A}t} + \int_{0}^{t} e^{-\mathbf{A}(t - t')}\mathbf{B}d\mathbf{W}(t'). \label{eq:pll_sde_y}
\end{align}
Note that the \textit{``starting state''} of the \ac{OU} process $\mathbf{y}(0)$ is a vector of random variables.
Since the starting value is not significant for the statistical properties of the process, it is defined as a deterministic value $\mathbf{y}(0) \equiv 0$.
Vector $\mathbf{y}(t)$ consists of a vector of wiener processes $d\mathbf{W}(t)$.
Its increments are zero mean, independent, Gaussian-distributed random variables.
Therefore, the vector of state variables is also zero mean.

The stochastic time shift at the output of the \ac{PFD} is
\begin{align}
    \beta(t) &= \alpha_\text{PLL}(t) - \alpha_\text{REF}(t)\label{eq:pll_output_process} \\
             &= \alpha_\text{PLL}(t) - \sqrt{c_\text{REF}} \int_{0}^{t} \xi_\text{REF}(t')dt'.\label{eq:pll_output_process2}
\end{align}
Note that here the assumption is that the reference oscillator is a free-running oscillator (equivalent to the model established previously), characterized by the oscillator constant $c_\text{REF}$.
The stochastic time shift at the output of the \ac{PLL} is
\begin{align}
    \alpha_\text{PLL}(t) = \sqrt{c_\text{CTL}} \int_{0}^{t} \gamma(t')dt' + \sqrt{c_\text{VCO}} \int_{0}^{t} \xi_\text{VCO}(t')dt'.
    \label{eq:pll_output_process3}
\end{align}
Using (\ref{eq:pll_output_process2}) and (\ref{eq:pll_output_process3}), the differential in time for the process $\beta(t)$ can be expressed as
\begin{align}
     \frac{d\beta(t)}{dt} &= \sqrt{c_\text{CTL}}~\gamma(t) + \sqrt{c_\text{VCO}}~\xi_\text{VCO}(t) - \sqrt{c_\text{REF}}~\xi_\text{REF}(t)\nonumber\\
                          &= \sqrt{c_\text{CTL}}\left( -k_\text{PFD}\beta(t) \ast h_\text{LF}(t) \right)\nonumber\\
                          &+ \sqrt{c_\text{VCO}}~\xi_\text{VCO}(t) - \sqrt{c_\text{REF}}~\xi_\text{REF}(t).
\end{align}
    \section{Autocorrelation Function of Phase-Locked Loop Output Process}
\label{appendix:acf_pll_out}

In order to find an expression for $R_{\alpha,\alpha}$, the \ac{ACF} $R_{\beta,\beta}$ and \ac{CCF}'s $R_{\alpha,\beta}, R_{\beta,\alpha}$ have to be identified.
The \ac{ACF} for the \ac{OU}-process $\beta(t)$ is found as
\begin{align}
    R_{\beta,\beta}(t, t + \tau) = \sum_{i=1}^{p} \nu_i \sum_{j=1}^{p} \left( e^{-\lambda_i |\tau|} - e^{- (\lambda_i t + \lambda_j (t + \tau))} \right).
\end{align}
The variables $\lambda_i, \lambda_j, \nu_i$ are introduced in \cite{mehrotra1031966}.
In the considered configuration of the \ac{PLL} they are given as
\begin{align}
    \lambda_i &= \lambda_j = \lambda = 2 \pi f_\text{PLL}\\
    \nu_i &= \nu = \frac{c_\text{VCO} + c_\text{REF}}{4 \pi f_\text{PLL}}.
\end{align}
The \ac{ACF} over time becomes asymptotically \cite[cf. (11)]{mehrotra1031966}
\begin{align}
    \lim_{t\to\infty} s_{\beta,\beta}(t, t + \tau) = \sum_{i=1}^{p} \nu_i e^{-\lambda_i |\tau|}.
\end{align}
The \ac{CCF} component can be calculated using (\ref{eq:pll_sde_y}), \eqref{eq:pll_sde_beta}. 
The integral is solved and mathematical transformations are applied to find
\begin{align}
    R_{\alpha,\beta}(t, t+\tau) = \sqrt{c_\text{REF}}~ \sum_{i=1}^{p} \mu_i \left( e^{-\lambda_i |\tau|} - e^{-\lambda_i (t + \tau)}\right),
\end{align}
and analogously
\begin{align}
    R_{\beta, \alpha}(t, t+\tau) = \sqrt{c_\text{REF}}~ \sum_{i=1}^{p} \mu_i \left( 1 - e^{-\lambda_i t} \right).
\end{align}
Using the notation in \cite{mehrotra1031966} a constant factor is introduced which in our scenario equates to 
\begin{align}
    \mu_i = \mu = - \frac{c_\text{REF}}{2 \pi f_\text{PLL}}.
\end{align}
Identical to \cite[(10)]{mehrotra1031966}, the sum of the \ac{CCF}'s asymptotically approaches
\begin{align}
        \lim_{t\to\infty} [ R_{\alpha,\beta} &+ R_{\beta,\alpha} ] = \sum_{i=1}^{p} \mu_i \left(  e^{-\lambda_i |\tau|} + 1 \right).
\end{align}
Finally the \ac{ACF} of $\alpha_\text{PLL}$ at the \ac{PLL} output can be expressed as
\begin{align}
    R_{\alpha,\alpha} (t, t + \tau) &= \sum_{i=1}^{p} \nu_i \sum_{j=1}^{p} \left[ e^{-\lambda_i |\tau|} - e^{-(\lambda_i t + \lambda_j(t+\tau))} \right] \nonumber\\
    &+ \sum_{i=1}^{p} \mu_i \left[  e^{-\lambda_i |\tau|} - e^{-\lambda_i (t + \tau)} + 1  - e^{-\lambda_i t }\right]\nonumber\\
    &+c_\text{REF} t.
\end{align}
    \section{Power Spectrum Density of Phase-Locked Loop}
\label{appendix:psd_pll}

Equivalent to \eqref{eq:acf} an expression that includes a difference term $\Delta\alpha = \alpha_\text{PLL}(t) - \alpha_\text{PLL}(t + \tau)$ and a sum term $\Sigma\alpha = \alpha_\text{PLL}(t) + \alpha_\text{PLL}(t + \tau)$ can be obtained.
The terms of the expectation of difference and sum are again replaced by their characteristic functions $\Psi_{\Delta\alpha}, \Psi_{\Sigma,\alpha}$ (cf. \eqref{eq:psi_delta}, \eqref{eq:psi_sigma}).
Similar to \eqref{eq:var_diff}, \eqref{eq:var_sum} the variance for the difference and sum processes has to be calculated.
With \eqref{eq:acf_pll} the variance for the difference process can be written as
\begin{align}
    &\text{Var}(\Delta\alpha) = \text{Var} (\alpha_\text{PLL}(t) - \alpha_\text{PLL}(t + \tau))\nonumber\\
    &= \text{Var} (\alpha_\text{PLL}(t))  - 2 R_{\alpha,\alpha} (t, t + \tau) + \text{Var} (\alpha_\text{PLL}(t + \tau))\nonumber\\
    &= {\sum_{i=1}^{p} \nu_i \sum_{j=1}^{p} \left[ 2 - e^{-(\lambda_i + \lambda_j)t} - e^{-(\lambda_i + \lambda_j)(t + \tau)} - 2 e^{-\lambda_i |\tau|} \right.}\nonumber\\
    &{\left. + 2e^{-(\lambda_i t + \lambda_j(t + \tau))} \right]} + {2 \sum_{i=1}^{p} \mu_i \left[ 1 - e^{-\lambda_i |\tau|} \right]} + { c_\text{REF} |\tau|}.
\end{align}
Asymptotically over time the variance for $\Delta\alpha$ becomes
\begin{align}
    \lim_{t\to\infty} \text{Var}(\Delta\alpha) = 2 \sum_{i=1}^{p} (\nu_i + \mu_i) \left[ 1 - e^{-\lambda_i |\tau|} \right] + c_\text{REF} |\tau|.
\end{align}
In a similar fashion, the variance of the sum process (cf. \eqref{eq:var_sum}) is found
\begin{align}
    &\text{Var}(\Sigma\alpha) = \text{Var} (\alpha_\text{PLL}(t) + \alpha_\text{PLL}(t + \tau))\nonumber\\
    &= \text{Var} (\alpha_\text{PLL}(t))  + 2 R_{\alpha,\alpha} (t, t + \tau) + \text{Var} (\alpha_\text{PLL}(t + \tau))\nonumber\\
    &= \hdots + c_\text{REF} (4t + \tau).
\end{align}
Asymptotically over time the variance for $\Sigma\alpha$ becomes
\begin{align}
    \lim_{t\to\infty} \text{Var}(\Sigma\alpha) = \infty,
\end{align}
and therefore $\Psi_{\Sigma\alpha} \rightarrow 0$.
Then the asymptotic \ac{ACF} of the $\alpha_\text{PLL}$ process can be written as
\begin{align}
    &\lim_{t\to\infty} R_{x,x}(t, t + \tau) = \sum_{i=\pm 1} \left| \underline{X}_i \right|^2 e^{ji 2 \pi f_0 \tau} \Psi_{\Delta\alpha}\nonumber\\
    &+ \underbrace{\underline{X}_i \underline{X}_{-i}^{*} e^{ji 2 \pi f_0 (2t + \tau)} \Psi_{\Sigma\alpha}}_{=0,~\Psi_{\Sigma\alpha} \rightarrow 0} \nonumber\\
    &= \sum_{i=\pm 1} \left| \underline{X}_i \right|^2 e^{ji 2 \pi f_0 \tau}\nonumber\\
    &\cdot e^{- (2 \pi f_0)^2 \left[ 
    \sum_{k=1}^{p} (\nu_k + \mu_k) \left[ 1 - e^{-\lambda_k |\tau|} \right] + 0.5 c_\text{REF} |\tau| \right]}.
\end{align}
In order to make calculating the Fourier transform of the \ac{ACF} easier, mathematical transformations are applied to receive
\begin{align}
        &\lim_{t\to\infty} R_{x,x}(t, t + \tau) = \sum_{i=\pm 1} \left| \underline{X}_i \right|^2 e^{ji 2 \pi f_0 \tau} \nonumber\\
        &\cdot \left[ \sum_{n_1=0}^{\infty} \hdots \sum_{n_p=0}^{\infty} \left( \eta_{n_k} \cdot e^{- \kappa_{n_k} |\tau|} \right) \right]
\end{align}
The parameters $\eta_{n_k}, \kappa_{n_k}$ are introduced to allow for more compact notation and represent
\begin{align}
    \eta_{n_k} &= e^{- (2 \pi f_0)^2 \sum_{k=1}^{p} (\nu_k + \mu_k)} \prod_{k=1}^{p} \frac{{\left( ( 2 \pi f_0)^2 (\nu_k + \mu_k) \right)}^{n_k}}{n_k!}\\
    \kappa_{n_k} &= 0.5 ( 2 \pi f_0)^2 c_\text{REF} + \sum_{k=1}^{p} n_k \lambda_k
\end{align}
In the specific scenario considered for $p=1$ the parameters simplify to
\begin{align}
    \eta_{n} &= \frac{1}{n!} e^{- \frac{\pi f_\text{0}^2}{f_\text{PLL}} \left( c_\text{VCO} - c_\text{REF} \right)} {\left( \frac{\pi f_\text{0}^2}{f_\text{PLL}} \left( c_\text{VCO} - c_\text{REF} \right) \right)}^{n}\\
    \kappa_{n} &= 2 \pi \left( \pi f_\text{0}^2 c_\text{REF} + n f_\text{PLL} \right).
\end{align}
The \ac{PSD} of the \ac{PLL} output is 
\begin{align}
    &S_{x,x}(f) = \mathfrak{F}_\tau \Bigl\{ R_{x,x} (\tau) \Bigr\}\nonumber\\
    &= \sum_{i=\pm 1} \left| \underline{X}_i \right|^2 \delta(f - if_\text{0}) \ast \left[ \sum_{n=0}^{\infty} \eta_{n} \frac{2 \kappa_{n}}{\kappa_{n}^2 + f^2}  \right]\nonumber\\
    &= \sum_{i=\pm 1} \left| \underline{X}_i \right|^2 \sum_{n=0}^{\infty} \eta_{n} \frac{2 \kappa_{n}}{\kappa_{n}^2 + (f - i f_\text{0})^2}. 
\end{align}

    \section{Phase Noise Model Parameters}
\label{appendix:pn_model_params}

Values for the phase noise model parameter estimates for the spectrum displayed in \figref{fig:pn_lin_regr} are given in the following.
$\SI{3}{dB}$ cut-off frequency of \ac{REF} and \ac{VCO}
\begin{align*}
    \hat{f}_\text{3dB,REF} = \SI{0.58}{Hz},~\hat{f}_\text{3dB,VCO} = \SI{630}{Hz}.
\end{align*}
Oscillator constant for \ac{REF} and \ac{VCO}
\begin{align*}
    \hat{c}_\text{REF} = \SI{4.58e-20}{s},~\hat{c}_\text{VCO} = \SI{5.01e-17}{s}
\end{align*}
Transition interval and noise floor power level
\begin{align*}
    \hat{\mathfrak{L}}_\text{TR} = -107.9~\text{dBc/Hz},~\hat{\mathfrak{L}}_\text{NF} = -133.7~\text{dBc/Hz}.
\end{align*}
Transition interval start frequency, \ac{PLL} bandwidth and noise floor start frequency
\begin{align*}
    \hat{f}_\text{TR} = \SI{1865.7}{Hz},~\hat{f}_\text{PLL} = \SI{197.9}{kHz},~\hat{f}_\text{NF} = \SI{1439.8}{kHz}.
\end{align*}

Parameter estimates derived from the whole dataset for the two daughterboard types are presented in \tabref{table:parameter_estimates}.

\begin{table}[ht]
    \centering
\begin{center}
        \resizebox{0.6\linewidth}{!}{
\begin{tabular}{| l | l |}
 \hline
 Parameter & CBX \\
 \hline
 $\hat{f}_\text{3dB,REF}$ & $0.5570 \pm 0.0249 ~ \text{Hz}$  \\ 
 $\hat{c}_\text{REF}$ & $ (4.432 \pm 0.1978) \cdot10^{-20} ~ \text{s}$ \\ 
 $\hat{f}_\text{3dB,VCO}$ & $193.4 \pm 16.69 ~ \text{Hz}$ \\ 
 $\hat{c}_\text{VCO}$ & $ (1.539 \pm 0.1328)\cdot10^{-17} ~ \text{s}$ \\ 
 $\hat{\mathfrak{L}}_\text{TR}$ & $-91.9 \pm 1.735 ~ \text{dBc}$\\ 
 $\hat{f}_\text{TR}$ & $538.7 \pm 64.25~ \text{Hz}$ \\ 
 $\hat{f}_\text{PLL}$ & $26.6 \pm 3.8 ~ \text{kHz}$ \\ 
 $\hat{\mathfrak{L}}_\text{NF}$ & $-144.4 \pm 0.2197 ~ \text{dBc}$\\ 
 $\hat{f}_\text{NF}$ & $1487 \pm 92.89~ \text{kHz}$\\ 
 \hline
 Parameter & UBX \\
 \hline
 $\hat{f}_\text{3dB,REF}$ & $ 0.5853 \pm 0.0503 ~ \text{Hz}$\\
 $\hat{c}_\text{REF}$ & $ (4.658 \pm 0.4005)\cdot10^{-20} ~ \text{s}$\\
 $\hat{f}_\text{3dB,VCO}$ & $537.6 \pm 63.26 ~ \text{Hz}$ \\
 $\hat{c}_\text{VCO}$ & $ (4.278 \pm 0.5034)\cdot10^{-17} ~ \text{s}$ \\
 $\hat{\mathfrak{L}}_\text{TR}$  & $-107.8 \pm 0.7843 ~ \text{dBc}$ \\
 $\hat{f}_\text{TR}$ & $1872.1 \pm 119.61~ \text{Hz}$ \\
 $\hat{f}_\text{PLL}$ & $177.3 \pm 20.04~ \text{kHz}$ \\
 $\hat{\mathfrak{L}}_\text{NF}$ & $-134.0 \pm 0.1847~ \text{dBc}$ \\
 $\hat{f}_\text{NF}$ & $1319 \pm 106.20~ \text{MHz}$ \\
 \hline
\end{tabular}
}
\end{center}
\caption{Parameter estimates for different daughterboard types given as $\mu \pm \sigma$} 
\label{table:parameter_estimates}
\end{table}
	\bibliographystyle{IEEEtran}
	\bibliography{references}
\end{document}